\documentclass[referee]{raa}

\usepackage{graphicx,times}
\usepackage{natbib}
\usepackage{amssymb,amsmath}
\usepackage{siunitx}
\usepackage{float}
\bibpunct{(}{)}{;}{a}{}{,}
\usepackage[utf8]{inputenc}
\usepackage[pagebackref=true]{hyperref}
\usepackage{orcidlink}
\usepackage{color}

\newcommand{\citeg}[1]{\citep[e.g.,][]{#1}}
\renewcommand{\abstractname}{Abstract} 
\renewcommand{\figurename}{Fig.} 
\DeclareSIUnit\angstrom{\text {Å}}

\begin{document}

\title{Distance estimation of the high-velocity cloud Anti-Center Shell}

 \volnopage{ {\bf 20XX} Vol.\ {\bf X} No. {\bf XX}, 000--000}
   \setcounter{page}{1}

   \author{
   Yu-Ting Wang \orcidlink{009-0007-8574-0890}, \inst{1,2} 
   Chao Liu \orcidlink{0000-0002-1802-6917}, \inst{1,2,3}
   Zhi-Yu Zhang \orcidlink{0000-0002-7299-2876} \inst{4,5}}

   \institute{
   Key Laboratory of Space Astronomy and Technology, National Astronomical Observatories, CAS, Beijing 100101, People’s Republic of China; {\it liuchao@nao.cas.cn}\\
    \and
    University of Chinese Academy of Sciences, Beijing 100049, People's Republic of China\\
    \and
    Zhe jiang Lab, Hangzhou, Zhejiang Provence, 31121, China\\
    \and
    School of Astronomy and Space Science, Nanjing University, Nanjing 210023, People's Republic of China\\
    \and
    Key Laboratory of Modern Astronomy and Astrophysics (Nanjing University), Ministry of Education, Nanjing 210023, People's Republic of China\\
\vs \no
   {\small Received 20XX Month Day; accepted 20XX Month Day}
}

\abstract{
High-velocity clouds (HVCs) are interstellar gas clouds whose velocities are incompatible with Galactic rotation. 
Since the first discovery of HVCs in 1963, their origins have been debated for decades but are still not settled down, because of the lack of vital parameters of HVCs, e.g., the distance. 
In this work, we determined the distance to the high-velocity cloud, namely the Anti-Center Shell (ACS). 
We trace the ACS with extinction derived from K-giant stars with known distances and with the diffuse interstellar band (DIB) feature at 5780~\AA~fitted on spectra of O- and B-type stars with distance.
As a result, we provide a lower limit distance of ACS as $\sim8$~kpc, which extends the lower limit outward by approximately 4~kpc compared to previous work. A byproduct of the DIB method is that we detected a bar-shaped structure with a unusually high positive
line-of-sight velocity. Its shape extends along the $(l,b)=(155, -5)^{\circ}$ sight-line and shows a slightly increasing trend in equivalent width and velocity as the distance increases. } 


\authorrunning{Wang et al.}            
\titlerunning{Distance estimation of Anti-Center Shell}  
\maketitle

%
\section{Introduction}           
\label{sect:intro}

High-velocity clouds (HVCs) are neutral or ionized gas clouds in the vicinity
of the Milky Way that are characterized by velocities incompatible with
Galactic differential rotation \citep{wakker1997}. Traditionally, astronomers
selected out HVCs by applying a crude and invariable cut of line-of-sight
velocity in the Local Standard of Rest (LSR) reference framework such as
$|v_{\rm{LSR}}|\gtrsim 90~\rm{km} \cdot \rm{s}^{-1}$.  After that, an improved
definition was proposed by \citet{wakker1991-distribution} who introduced a
so-called deviation velocity $v_{\rm{dev}}=50~\rm{km} \cdot \rm{s}^{-1}$ to
characterize HVCs as deviating by a fixed velocity separation from the
maximally permissible velocity of Galactic disc in a given direction.
Recently, \citet{westmeier2018-new} assumed a simple, cylindrical model of the
Galactic disc with a disc radius of 20~kpc, a disc height of 5~kpc, and a
deviation speed of $v_{\rm{dev}}=70~\rm{km} \cdot \rm{s}^{-1}$ to produce
velocity masks to filter out disk component based on the all-sky \ion{H}{i} $4\pi$ survey \citep[HI4PI;][]{HI4PI2016},
a database for Galactic \ion{H}{i} emission.  The map of \citet{westmeier2018-new}
offers a comprehensive and high-resolution vision of high-velocity clouds
and thus becomes the cornerstone of this field.

Apart from \ion{H}{i} 21-cm observations, another observational technique for
high-velocity clouds that deserves equal attention is the absorption-line
method, by which HVCs can be reached by absorption towards bright background
halo stars or quasars, mostly at ultraviolet wavelengths  \citep{smoker2011-distance,marasco2022-intermediate}. While 21-cm \ion{H}{i}
observations provide us a uniform all-sky HVC map \citep{westmeier2018-new},
the absorption-line method, on the other hand, can probe much lower
hydrogen column density and detect ionized materials in the high-velocity
clouds \citep[e.g.,][]{sembach2000-far,sembach2003-highly,lehner2001-hst}. 


Despite that nearly six decades have passed since the initial detection of the high-velocity cloud in 1963 \citep{muller1963-hydrogene}, the enigmatic origin of them persists as an unresolved problem. Based on current observations, high-velocity clouds require at least three formation mechanisms: one for the Magellanic Stream and its associated clouds, one for the Outer Arm Extension, and one for the rest of the high-velocity clouds. 
Currently, existing hypotheses can be roughly categorized into four groups according to the locations of HVCs:

\begin{itemize}
    \item[1)] The HVCs are situated in the solar neighborhood, indicating that they are originated from nearby supernova explosions \citep[e.g.,][]{heiles1979};
    \item[2)] The HVCs are gas at the disk-halo interface, indicating the transfer of gas between these two regions \citep[e.g.,][]{marasco2022-intermediate};
    \item[3)] The HVCs are situated in the Outer Galaxy, e.g., as the polar ring wrapping our Milky Way or as gas stripped or ejected from dwarf galaxies \citep[e.g.,][]{davies1972};
    \item[4)] The HVCs stem from an even more distant place, as gas from the Local Group or as infalling intergalactic matter \citep[e.g.,][]{Verschuur1969}.
\end{itemize}

Intriguing as the origin theories are, it is vital to pin down the fundamental properties (e.g., distance, metallicity, and 3D kinematics) of high-velocity clouds if we want to figure out the detailed origin mechanisms. Distance measurement is critical since it helps to distinguish which categories the birth of HVCs might fall into, as well as to quantify their fundamental physical properties, several of which directly scale with the distance (e.g., total mass, size, density, pressure, \citealt{wakker1997}). 
However, distance determination for high-velocity clouds is challenging. 
The most used method is the absorption-line method \citep{schwarz1995-distance}. Absorption line features produced by HVCs can be explored with spectroscopic observations of background objects with known distances, leading us to metallicity, velocity, and constraints of distance (upper limit and lower limit). 
Up till now, a considerable volume of work using the absorption-line method has been published \citeg{smoker2011-distance,schwarz1995-distance,ryans1997-distance,wakker2001-distances,thom2006-galactic,thom2008-accurate,wakker2007-distances,wakker2008-distances,lehner2010-origin,smoker2006-caii,tripp2012-21,peek2016-first,lehner2022-inter}. 
However, the absorption-line method has trouble giving precise distance to the high-velocity clouds because it suffers from the limitation of sparsely-distributed tracer stars \citep{marasco2022-intermediate} due to the deficiency of ultraviolet band spectra.

In this paper, we intend to use interstellar extinction to trace the high-velocity cloud Anti-Center Shell (ACS) and constrain its distance. When starlight penetrates gas, dust, etc., its energy will be attenuated by absorption or scattering, called interstellar extinction. Generally, the atomic cloud regions have higher densities than the ambient medium. Consequently, the extinction in the HVC sight line would increase more than in the surroundings. The location of the extinction excess will reveal our distance to the cloud. \citet{yan2019-distance,yan2021-improved,zucker2019-alarge,wang2020-distances} have all successfully measured the distance to various types of objects, such as molecular clouds, supernova remnants (SNRs), etc., using the extinction method. Numerous and extensively distributed K-giants in the Milky Way (MW), which are our extinction tracers, can be helpful to derive much more precise distances. Hence, they can be used to probe down to the inner three-dimensional structure of the ACS with the extinction method. Meanwhile, to take advantage of the velocity information in the \ion{H}{i} 21-cm emission observation, we take Diffuse Interstellar Band (DIB) feature as an auxiliary tracer to verify the position of the very step-like pattern in the distance-extinction plot independently caused by ACS. 

The paper is structured as follows. In Section \ref{sec:data}, we describe the sample we used and our filtering strategy. Section \ref{sec:method} describes how we utilized extinction and Diffuse Interstellar Bands (DIBs) to locate the high-velocity cloud Anti-Center Shell (ACS). Section \ref{sec:results} discuss the main results from above methods, in which we provide the lower distance limit of ACS and report a newly-discovered positive-velocity high-velocity DIB clump. We summarize and discuss our results in Section \ref{sec:conclusion}.

\section{Data Collection} \label{sec:data}

\subsection{Cloud properties} \label{sec:Sample} 

The Anti-Center Shell (ACS) is a shell-like low-latitude HVC located in the direction opposite to the Galactic center from our solar system \citep{westmeier2018-new}. This shell has a $\sim 40^{\circ}$ diameter and a mean line-of-sight velocity in the Local Standard of Rest of $\sim -100~\rm{km} \cdot \rm{s}^{-1}$. The ACS is chosen as our target because: 1) It is located in the anti-center region with low Galactic latitude, making it plausible to be found within 6~kpc from the Sun, where \textsl{Gaia} distance data stays valid; 2) Previous work suggested that ACS might require independent origin theory \citep{wakker1997}. Distance measurement can constrain its way of formation and provide insights into its interactions with the surrounding environment.

\subsection{Spectroscopically-identified parameters of K-giants from LAMOST Data Release 5} \label{sec:lamost data}

We employed interstellar extinction to illustrate the spatial distribution of
the ACS since dust and gas particles in the cloud are both extinction
generators \citep{draine2003-inter}. K-giant stars are commonly-used tracers in extinction/reddening studies because of their brightness and abundance,
making them substantially observed in the MW \citep{xue2016-precise}.

The Guoshoujing Telescope (LAMOST) is a 4-meter special quasi-meridian
reflecting Schmidt telescope located at Xinglong Station of National
Astronomical Observatory of China \citep{cui2012}. It recently completed its twelfth data release and enabled global access to its tenth release of data. Till today, it has released an unprecedentedly large amount of low- and mid- resolution stellar spectra of over 17,000,000.\footnote{The information can be found on their official website: \url{https://www.lamost.org/public/?locale=en}}

\citet{xu2020-exploring} compiled a catalogue of K-type giant stars using
LAMOST DR5, which aligns well with the objectives of this study.  Atmospheric
parameters (e.g., effective temperature $T_{\mathrm{eff}}$, and surface gravity
$\mathrm{log}g$) determined using Stellar LAbel Machine (SLAM)
\citep{zhang2020-deriving} can be extracted from this catalog. According to
\citet{xu2020-exploring}, K-giant stars in LAMOST were selected following the
criteria presented in \citet{liu2014-giant}: 

\begin{align}
  \begin{array}{l}
  (4000<T_{\mathrm{eff}}<4600~\mathrm{K},\quad \mathrm{log} g<3.5)
  \lor 
  (4600<T_{\mathrm{eff}}<5600~\mathrm{K},\quad \mathrm{log} g<4.0)
  \end{array} \label{eq:1}
\end{align}

Here, $T_{\rm{eff}}$ and $\mathrm{log}g$ stand for effective temperature and
surface gravity, respectively. They also conducted a metallicity cut of
$\mathrm{[M/H]}>-1.2$ to avoid halo stars, which nicely favors our work since this will alleviate the metallicity's influence on color.

We additionally adopt a quality cut on the g-band signal-to-noise ratio \textit{snrg} to ensure the reliability of the effective temperature
$T_{\mathrm{eff}}$:

\begin{equation}
 \textit{snrg}>15.
\end{equation}\label{eq:2}

To ensure a reliable statistic analysis, the number of sources needs to be
large enough in each 100-K wide effective temperature bin. Therefore, we only
kept stars in the temperature range:

\begin{equation}
4000<T_{\mathrm{eff}}<5300~\mathrm{K}.
\label{eq:3}
\end{equation}

\subsection{2MASS All Sky Data Release: Point Source Catalog} \label{sec:2mass}

The Two Micron All Sky Survey (2MASS) is an all-sky survey using two
1.3~m telescopes located at Mount Hopkins, Arizona, and Cerro Tololo, Chile.
This survey encompasses three near-infrared passbands, namely $J$, $H$, and
$K_\mathrm{S}$. The photometry was verified to have approximately 0.02~mag
accuracy \citep{skrutskie2006-2mass}.

We adopted the $J$- and $K_\mathrm{S}$-band photometry from the 2MASS data to
derive the color index $J-K_\mathrm{S}$.  We made a cut on the photometric
quality flag \texttt{ph\_qual} to keep only the sources with the flag value
\texttt{A}. The value of \texttt{ph\_qual} is set based on
signal-to-noise ratio, measurement quality, and detection
statistics. Besides, a \texttt{cc\_flg} screening was also
included to remove sources probably contaminated by proximity. The precise
description of these flags can be retrieved at
\footnote{\url{https://irsa.ipac.caltech.edu/data/2MASS/docs/releases/allsky/doc/explsup.html}}.

\begin{align}
\begin{array}{l}
\texttt{ph\_qual}=``A*A"\\
\texttt{cc\_flg}=``0*0"
\end{array}
\label{eq:4}
\end{align}

Here, the asterisks mean that we did not put any constraint on the photometry
quality of the $H$-band data.

\subsection{Bailor-Jones' distance estimation based on \textit{Gaia} EDR3} \label{sec:bailor distance}

We adopted the distance estimated in the literature \citep{bailorjones2021-est}, which is based on the \textsl{Gaia} EDR3 Catalog. This work estimated distances for 1.47 billion objects in \textsl{Gaia} EDR3 using a Bayesian approach based on a prior built from a three-dimensional Galactic model that considered interstellar extinction and the magnitude limit of \textsl{Gaia}. Two types of distance are provided alongside the article. The geometric one only used \textsl{Gaia} EDR3 parallax and the geometric prior above. The second, photogeometric distance, added color and apparent magnitude to give extra constraints. We chose to use the geometric distance \texttt{rgeo} in this article to avoid inviting color and magnitude reliance. 

\citet{bailorjones2021-est} recommended using the various quality fields in the main \textsl{Gaia} catalog of EDR3 for quality filtering purposes, rather than using the flags in \citet{bailorjones2021-est} to select high-quality distance estimates. 
We used the following constraints in this study:

\begin{align}
\begin{array}{l}
\texttt{ruwe}<1.4,\\
\texttt{parallax\_over\_error}>5.
\end{array}
\label{eq:5}    
\end{align}

\subsection{Diffuse Interstellar Band (DIB) feature as the cloud tracer} \label{sec:DIB} 

\subsubsection{Catalog of 16,002 O- and B-type Stars from LAMOST} \label{sec:OB xiang}

The Diffuse Interstellar Bands (DIBs) are a large set of absorption features
that arise in interstellar gas \citep{herbig1995-diffuse}. In this work, we
focused only on the absorption at about \qty{5780}{\angstrom}, which has been
validated as highly-correlated with the intensity of interstellar reddening and
\ion{H}{i} 21-cm emission \citep{herbig1995-diffuse, lan2015-ex}. It is convenient to
extract the DIB $\lambda5780$ from the spectra of O- and B-type stars since
they do not show many other stellar features at the red end.

The catalog of 16,002 O- and B-type stars from LAMOST by \citet{xiang2021-data} is perfectly suited for this process, so we selected the O- and B-type stars in the ACS direction from their catalog and downloaded corresponding low-resolution spectra from LAMOST Data Release 8\footnote{\url{https://www.lamost.org/public/}}.

\section{Method}\label{sec:method}

\subsection{Extinction as the cloud tracer} \label{sec:method1} 

Because the attenuation of light by interstellar extinction is wavelength-dependent, which is more efficient in the bluer wavelengths, stellar lights are typically reddened when they penetrate through the interstellar medium \citep{draine2003-inter}. The interstellar extinction ($A_{\lambda}$) and reddening (characterized by the color excess $E(\lambda_1-\lambda_2)=A_{\lambda_1}-A_{\lambda_2}$) are therefore highly-correlated, and their relationship can be demonstrated by the extinction law \citep{wang2019-optical}. Given the definition of interstellar reddening, we calculated the color excess $E(J-K_\mathrm{S})$ of the K-giant stars with the equation:

\begin{equation}
E(J-K_\mathrm{S})=(J-K_\mathrm{S})-(J-K_\mathrm{S})_0,
\label{eq:6}
\end{equation}
where $(J-K_\mathrm{S})$ means the observational color, and $(J-K_\mathrm{S})_0$ is the intrinsic color of stars.

We determined the intrinsic color $(J-K_\mathrm{S})_0$ by assuming that the color of the bluest star at a specific effective temperature is the intrinsic color at this temperature. According to \citet{ducati2001-intrinsic,wang2014-universality,xue2016-precise}, for tracer stars distributing sufficiently broad in space, the bluest stars can be treated as a reference without extinction. In practice, we divided the K-giant stars into 100-K wide effective temperature bins and computed the median color of the top 5-percent bluest stars in each bin. After that, we fitted the $\left\{T_{\mathrm{eff}},\ (J-K_\mathrm{S})_0 \right\}_i$ relation with a cubic polynomial, obtaining an effective temperature - intrinsic color relation as in Equation \ref{eq:7}. With this relation, we can gain the reddening for every star with Equation \ref{eq:6}. 

\begin{figure}[h]
   \centering
  \includegraphics[scale=0.45]{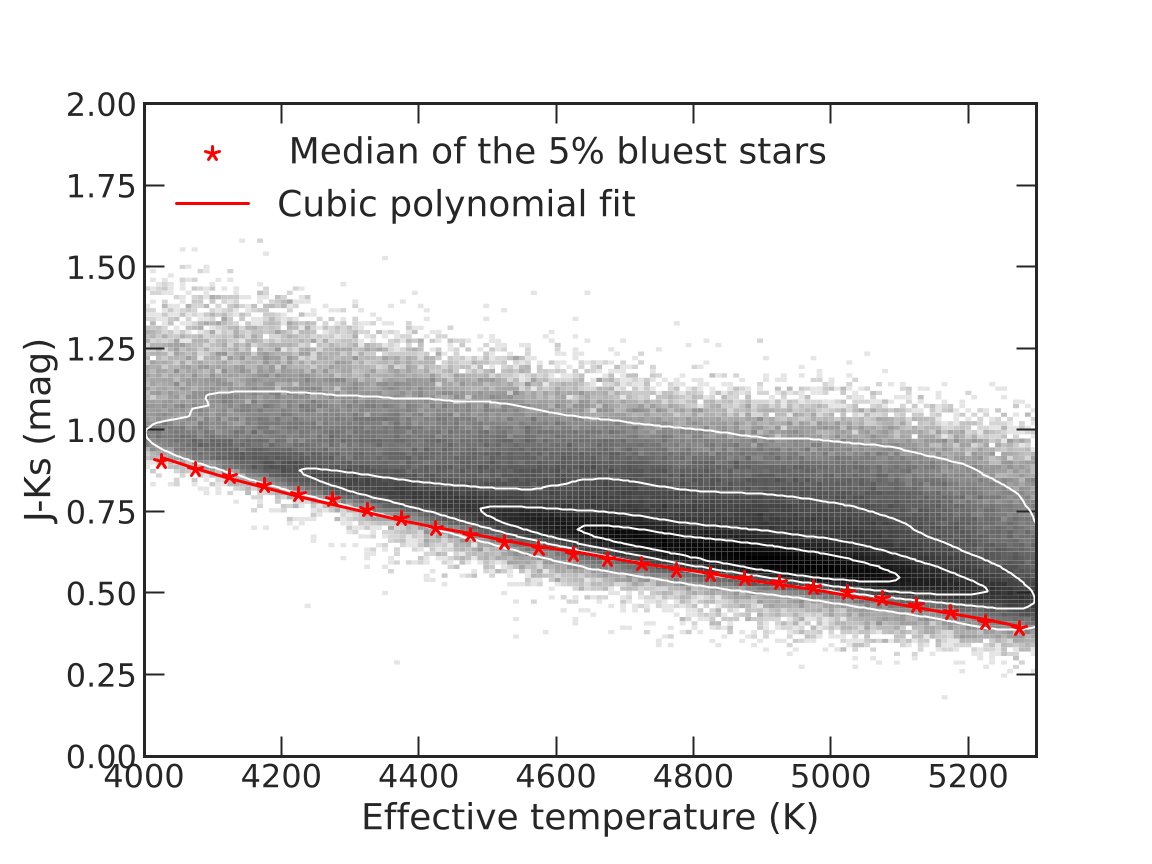}
   \caption{All the K-giant samples plotted on the $T_{\mathrm{eff}}-(J-K_\mathrm{S})$ diagram. The red asterisks denote the median color of the 5\% bluest stars in each $T_{\mathrm{eff}}$ bin. The red curve shows the cubic-polynomial fitting result to the red asterisks, which was considered as the derived $T_{\mathrm{eff}}-(J-K_\mathrm{S})_0$ relation.} 
   \label{fig:blueedge}
   \end{figure}

\begin{figure}[h]
\centering
\includegraphics[scale=0.65]{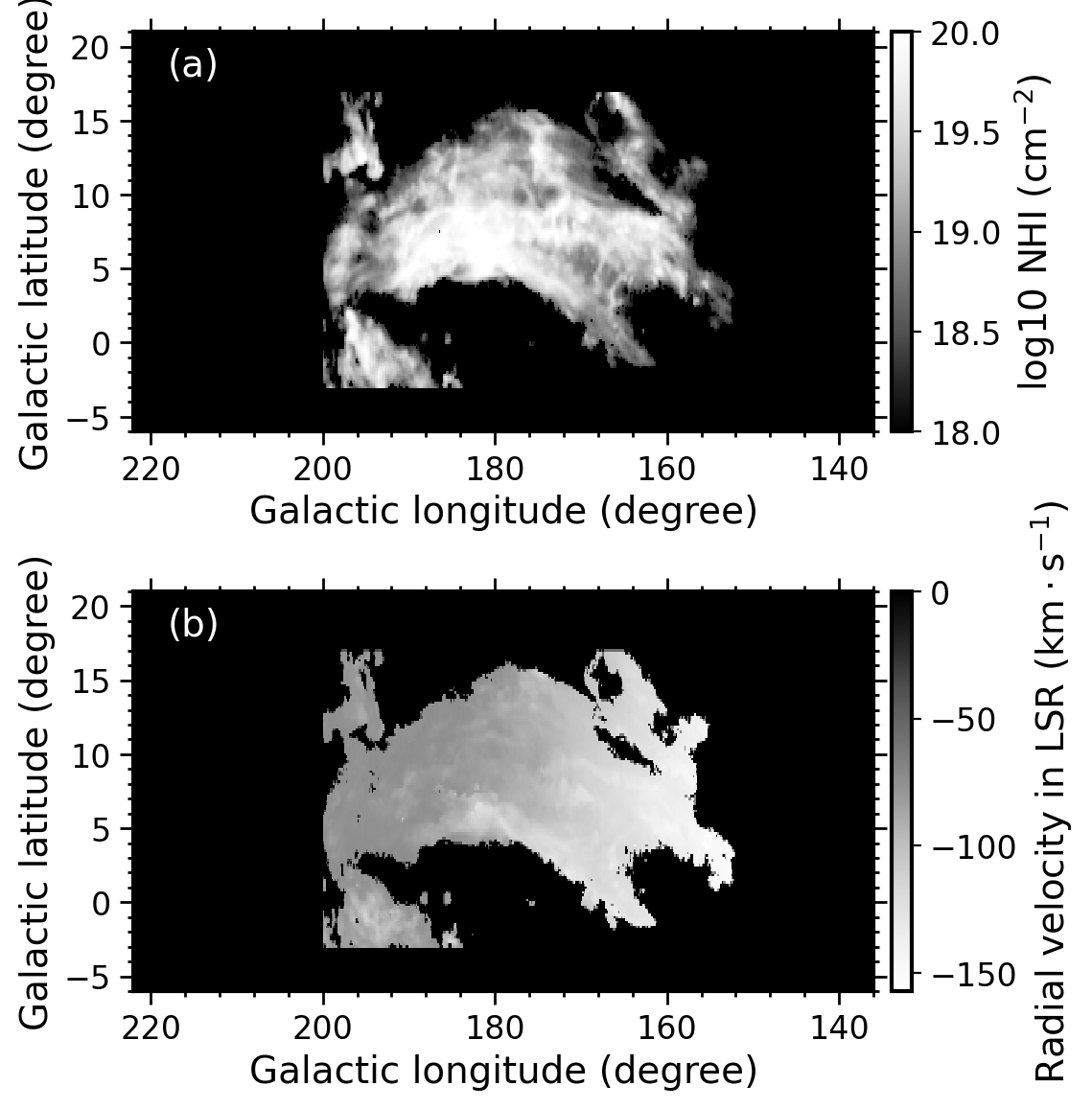}
\caption{\ion{H}{i} 21-cm emission map of Anti-center Shell (ACS) from HI4PI survey. The color map shows the column density (a) and mean radial velocity (b) distribution of neutral hydrogen at each sight line on the cloud.} 
\label{fig:cloud}
\end{figure}

To extract the extinction excess caused by ACS alone, we first sorted out tracer stars projecting inside the cloud profile and also those surrounding the cloud as a control group. By comparing the two, the cloud signal can be highlighted. In this part, we exploited the all-sky HVC map constructed by \citet{westmeier2018-new} to create an ACS mask. Stars located within the 21-cm \ion{H}{i} emission region of the ACS are assigned to the ``on-cloud'' group, while others to the ``off-cloud'' control group. The illustration of cloud profile is shown in Fig. \ref{fig:cloud}, color-coded by its logarithm \ion{H}{i} column density and radial velocity in LSR.

\begin{align}
(J-K_S)_0 = & -3.94\times10^{-11}T_{\rm{eff}}^3+6.94\times10^{-7}T_{\rm{eff}}^2 \notag \\
& -4.26\times10^{-3}T_{\rm{eff}}+9.36
\label{eq:7}
\end{align}

Now we have stars that distribute merely ``on'' the cloud, enabling us to trace the extinction variation along the sight line of ACS. The next step involves classifying these stars into distinct distance intervals. Here, we chose 200~pc as the bin size, a value that strikes a balance between maintaining acceptable measurement errors and ensuring noticeable resolution of the fine structure within the variation curve. Through shifting our temperature-intrinsic color relation line (also referred to as the ``bluest edge'' in the following text) to minimize the difference between the stars in each distance bin in the distance-color plot and the shifted bluest edge, we obtain the mean reddening suffered by these tracers, being the amount of displacement we applied to the bluest edge. The resulting Fig. \ref{fig:cloudE} illustrates the obtained extinction variation curve, with the error bars in the graph reflecting the comprehensive uncertainty including the uncertainty of the blue edge, the observational uncertainty of distance measurement, $J$-band and $K_{\mathrm{S}}$-band photometry, and the emerged Poison error when binning the data into sight line and distance bins. Detailed error estimation process for the extinction method is described in Appendix \ref{sec:error_est}.

The Anti-center Shell is a high-velocity cloud in the low-latitude region, which is uncommon among the HVC family \citep{westmeier2018-new}. This feature complicates the recognition process for the exact step-like pattern that ACS produces because the presence of the Galactic disk and the extensive dust in it will lead to a large amount of additional foreground/background extinction. As depicted in Fig. \ref{fig:cloudE}, there are several ``stairs'' as the distance increases, resulting from different sub-structures. Consequently, including a control group in the study is necessary. The control group encompasses a larger area beyond the cloud and is confined within Galactic longitude $\in \left[135,220\right]^{\circ}$ and Galactic latitude $\in \left[-5,25\right]^{\circ}$. To avoid excessive information loss caused by averaging over such a large region, we grouped the samples based on latitude. We split the on-cloud and off-cloud groups into low-latitude, medium-latitude, and high-latitude parts, respectively, and compared the derived extinction variation within the same latitudes, as illustrated in Fig. \ref{fig:Ecurve}. The major difference in overall growths of $E(J-K_S)$ among these curves originates from global extinction variation in the MW. Detailed analysis will be done in the Section \ref{sec:results}.

\subsection{Diffuse Interstellar Band (DIB) feature as the cloud tracer} \label{sec:method2}

From the spectra of O- and B-type stars, we can extract the line-center and equivalent width of the interstellar absorption feature at the location of each tracer by straightforward Gaussian fitting, and the continuum is estimated and deducted using the iterating method described in \citet{zhaohe2021-diffuse}. 

The iterative continuum fitting process from \citet{zhaohe2021-diffuse} began by fitting the $\lambda 5780$ local spectrum with a second-order polynomial and calculating the standard deviation of flux differences between individual pixels and the fit curve. Pixels above the curve were replaced by the corresponding points on the fitting curve if they exceeded $5\sigma$, while those below were replaced at $0.5\sigma$. The process was iteratively repeated 20 times using the remaining valid pixels and replacement points, with the final polynomial fit serving as the continuum for renormalization.

Line center can be converted into the line-of-sight velocity (also radial velocity above) of DIB $\lambda 5780$ carrier, while the absorption intensity, quantified by equivalent width, has been established by previous studies to exhibit a strong correlation with both the level of reddening and the 21-cm \ion{H}{i} emission \citep{herbig1995-diffuse, lan2015-ex}.
Consequently, this method not only allows us to construct a distance-reddening plot, which in this case takes the form of a distance-equivalent width plot, but also introduces an additional crucial dimension \textit{radial velocity}. The inclusion of velocity measurements enables us to identify signals associated with high-velocity clouds (HVCs) in the \ion{H}{i} observations.

After fitting the continuum and line center, we performed a series of quality cuts to retain measurements with reliable DIB $\lambda5780$ detection. First of all, we required the spectrum signal-to-noise ratio $\textit{snrg}>35$ to guarantee that the quality of our spectra was good enough for DIB $\lambda5780$ measurement. We also used the ratio of equivalent width value to equivalent width error to keep only $\texttt{EW5780/EW5780\_err}>4$ as ``detection'' of DIB $\lambda5780$ feature (otherwise, as ``non-detection''). Finally, we checked our data on the equivalent width - reddening  plot ($\mathrm{EW}-E(B-V)$), excluding the points that deviate explicitly from a linear relation. Specifically, we fitted the $\mathrm{EW}-E(B-V)$ distribution to a linear relation and calculated the standard deviation $\sigma$ of our data from this linear relation. Data points deviating more than $3\sigma$ from this relation were excluded.

To filter out DIB $\lambda5780$ detections with velocities incompatible with Galactic differential rotation (also referred to as high-velocity DIBs, or HV-DIBs later), we applied the method employed by \citet{westmeier2018-new} to screen out high-velocity clouds in \ion{H}{i} 21-cm emission database. We adopted the rotation curve by \citet{clemens1985-massachusetts} and projected this rotation velocity onto line-of-sight direction using Equation (3) in \citet{westmeier2018-new}:

\begin{figure}[h]
\centering
\includegraphics[scale=0.53]{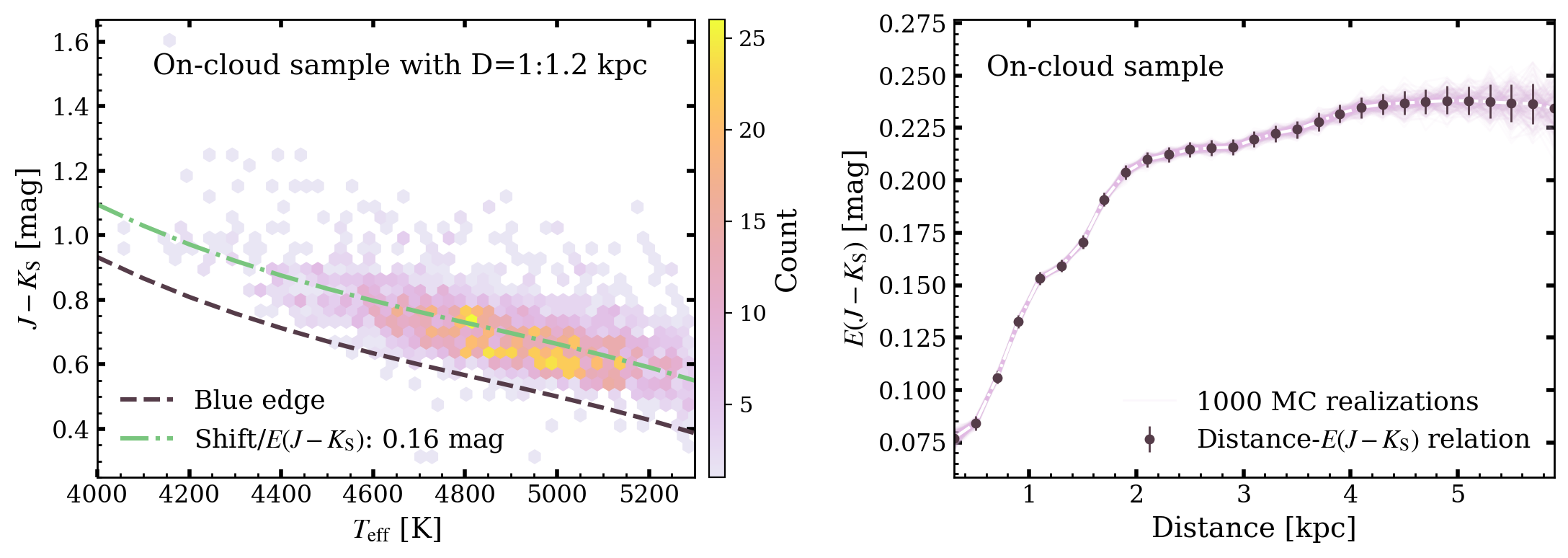}
\caption{(Left) Give an example of how the average reddening of one distance bin (from 1 to 1.2 kpc) is fitted. The hexagonal binning plot shows the number density distribution of the stars from 1 to 1.2 kpc in the temperature-color plane. The black dashed line illustrates the blue edge that we derived before. The green dotted-dashed line illustrates the best-fitted shift amount for the stars in the plot. (Right) We bin the on-cloud sample from 0 to 6 kpc with a distance width of 0.2 kpc. We resample the distance, temperature, and color $J-K_{\mathrm{S}}$  according to their observational uncertainties, and fit for the average $E(J-K_{\mathrm{S}})$ of each distance bin for 1000 runs. The MC results are plotted with pink solid lines. The median distance-$E(J-K_{\mathrm{S}})$ relation is plotted with black dots and white lines, with the uncertainty caused by the measurement of the blue edge already added in the error bars.}
\label{fig:cloudE}
\end{figure}

\begin{figure}[h]
\centering
\includegraphics[scale=0.6]{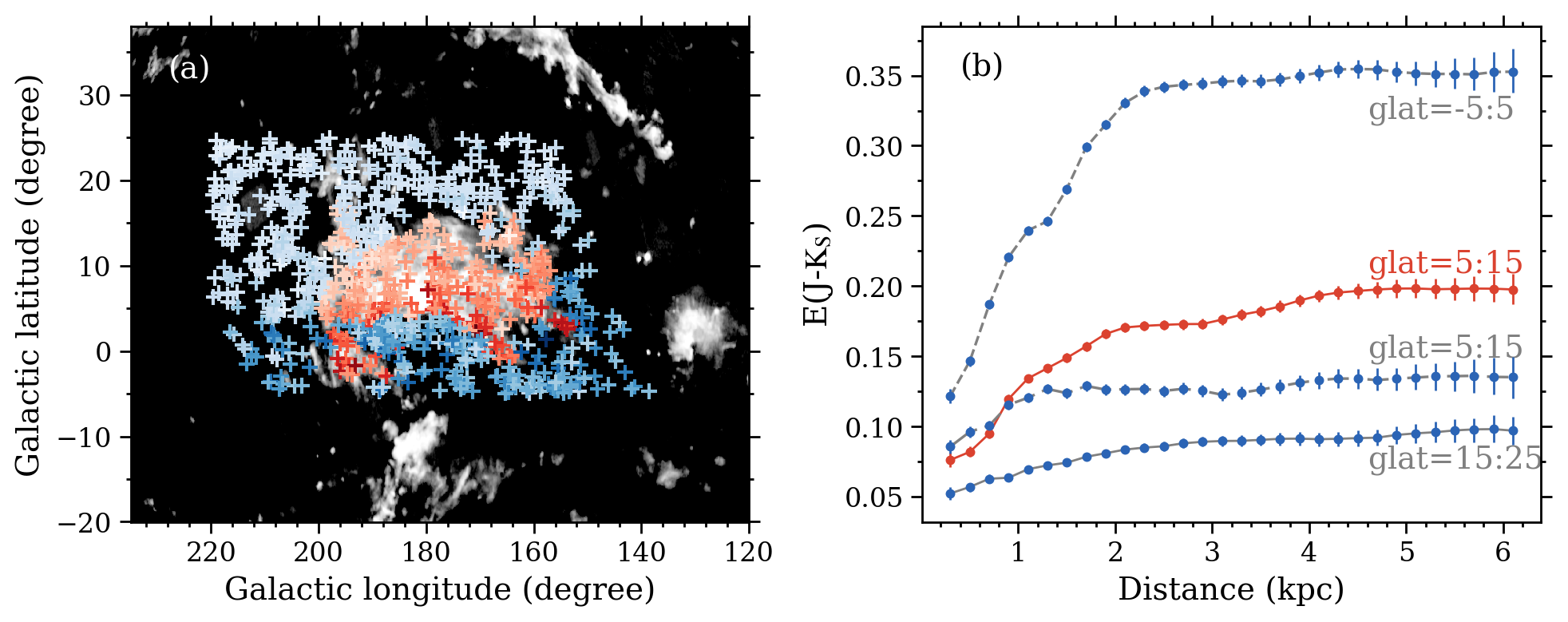}
\caption{(a) Illustration of the target regions of the four curves in Panel (b). The color intensity of red and blue represents the magnitude of reddening, respectively, for the stars on-cloud and off-cloud. (b) Mean reddening varies with distance along four different sight lines. The curve produced by stars on-cloud and at Galactic latitude $\in \left[5,15\right]^{\circ}$ is plotted with red crosses and a red solid line. The curves made by low-latitude, medium-latitude, and high-latitude off-cloud stars are drawn by blue crosses and gray dashed, solid, dash-dotted line, respectively.}
\label{fig:Ecurve}
\end{figure}

\begin{equation}
v_{\rm{proj}}=\left[v_{\rm{rot}}(r_{\rm{xy}}) \dfrac{r_{\odot}}{r_{\rm{xy}}} \right]\sin l \cdot \cos b
\label{eq:8}
\end{equation}
where $r_{\rm{xy}}$ equals the projection length of the Galactocentric distance onto xy-defined Galactic plane.
 
Assumed that the Galactic disk is a cylinder with $r_{\rm{max}}=20$~kpc and $|z_{\rm{max}}|=5$~kpc \citep[the same values as set by][]{westmeier2018-new}, we can calculate the distance range $\left[0, d_{\rm{max}}\right]$ encountered between the sun and the boundary of this cylindrical disc model at certain $(l,b)$.

Now we can obtain the projected velocity range of Milky Way rotation at any given $(l,b)$. We considered the DIB $\lambda 5780$ with a radial velocity deviating more than $70~\rm{km} \cdot \rm{s}^{-1}$ from this range as HV-DIB, and uncertainty in velocity determination is taken into account by Equation \ref{eq:9}. The result is shown in Fig. \ref{fig:HVDIBs}. The gray dots are all reliable DIB measurements selected with the above criteria. The shaded area represents the velocity range permitted by Galactic differential rotation. Normal velocity dispersion on the disk is included in the deviation velocity $v_{\mathrm{dev}}=70~\rm{km} \cdot \rm{s}^{-1}$. High-velocity DIB measurements, i.e., DIBs with anomaly velocity, are drawn with asterisks and are color-coded by their equivalent widths. 

\begin{equation}
v+\sigma_{v}<v_{\rm{proj,min}}-70 \ \mathrm{or} \ v-\sigma_{v}>v_{\rm{proj,max}}+70
\label{eq:9}
\end{equation}

It is easy to notice that there exist both negative velocity HV-DIBs and positive velocity ones in Fig. \ref{fig:HVDIBs}. ACS is a negative velocity HVC, which means it has a movement toward us, so we should primarily check these negative velocity HV-DIB points in space to see if they are statistically clustered to form a sub-structure like ACS. This will be discussed in the Section \ref{sec:results}.

\section{Results}\label{sec:results}
This part will discuss the main results from the extinction method and the DIB method in Section \ref{sec:method}.

\subsection{Bimodal distribution in distance - reddening scatter plot} \label{sec:res1}

In Fig. \ref{fig:cloudE}, the curve drawn with tracer stars projected on-cloud has several step-like patterns on it. However, it seems that none of them is solely displayed in the on-cloud plot according to Fig. \ref{fig:Ecurve}. The first two significant extinction steps at $1$~kpc and $2$~kpc are probably caused by the Local Arm and the Perseus Arm \citep{xu2023-what}. 

Since averaging reddening $E(J-K_S)$ in Fig. \ref{fig:Ecurve} may wipe out essential information, we again calculated individual reddening of all K-giant stars and produced a scatter plot in Fig. \ref{fig:bi}. There are two separated components in both plots: a high-extinction one probably stemming from sight lines going through the dusty Galactic disk, and a low-extinction component going through the higher-latitude low-density regime.

We manually separated these components in Fig. \ref{fig:bi} and illustrated their distribution in Galactic longitude-galactic latitude (glon-glat) space in Fig. \ref{fig:warp}. It is evident that the high-extinction components, both in the on-cloud and off-cloud groups, are predominantly concentrated in low-latitude regions, which are likely associated with the structure of the Galactic disk. In contrast, low-extinction ones are located at higher latitudes. This spatial distribution provides further support to our explanation regarding the origin of the bimodal pattern depicted in Fig. \ref{fig:bi}. We used a solid red line to specify the approximate boundary between the low-extinction and the high-extinction components in Fig. \ref{fig:warp}. The edge between the high and the low-extinction regime appears to be tilted, potentially indicating the presence of a warp in the interstellar dust distribution within our Galaxy \citep{freudenreich1994-dirbe}.

\begin{figure}[h]
\centering
\includegraphics[scale=0.45]{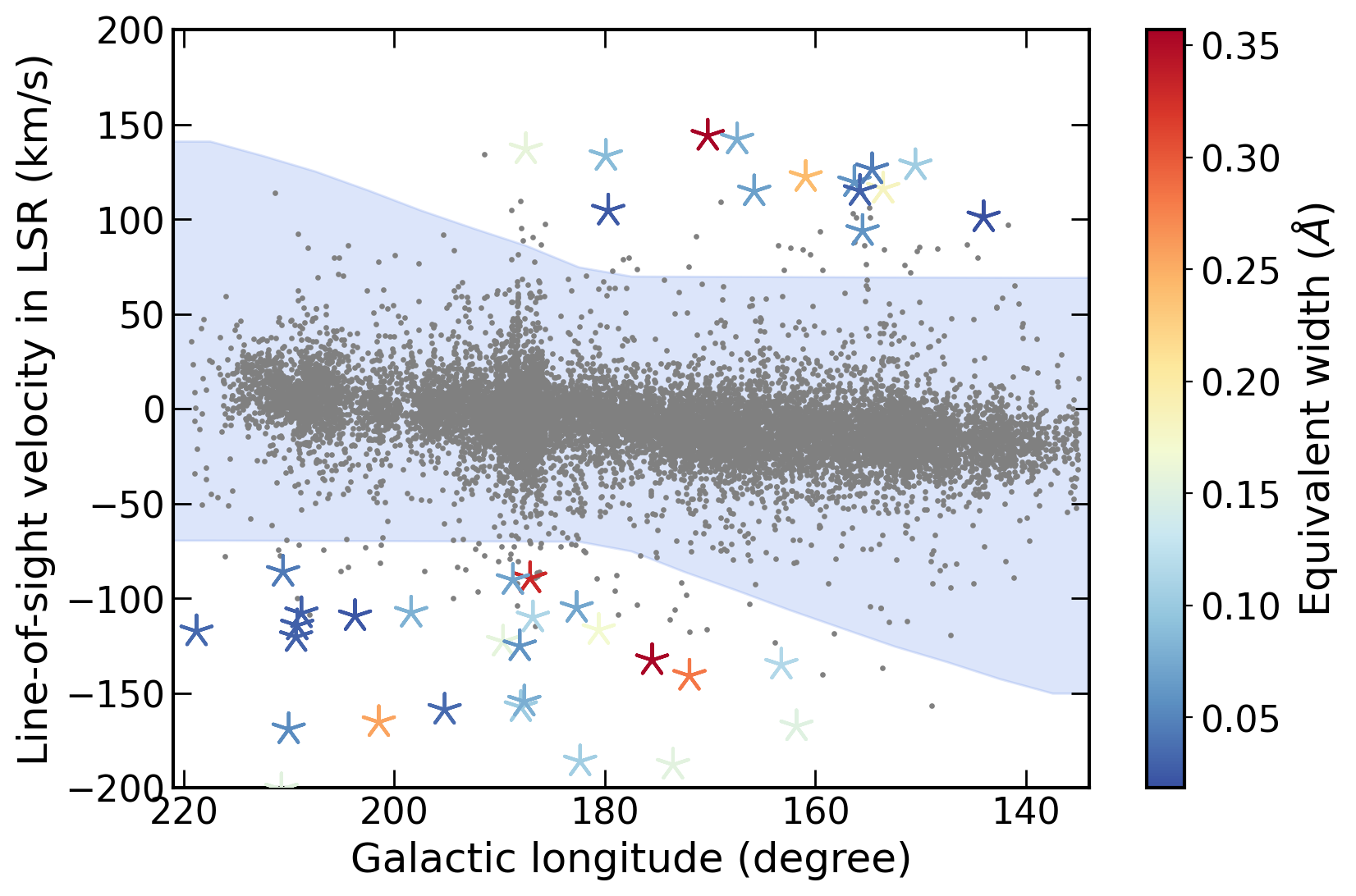}
\caption{DIB $\lambda 5780$ measurements in Galactic longitude - radial velocity plot. The gray dots are all reliable DIB measurements selected with criteria listed in Section \ref{sec:method2}. The shaded area implies the velocity range permitted by Galactic differential rotation. Normal velocity dispersion on the disk is included in the deviation velocity $v_{\mathrm{dev}}=70~\rm{km} \cdot \rm{s}^{-1}$. High-velocity DIB measurements, i.e., DIBs with anomaly velocity, are marked with asterisks and color-coded by their equivalent widths.} 
\label{fig:HVDIBs}
\end{figure}

We then regrouped these components according to their spatial distribution relative to the solid red line in Fig. \ref{fig:warp} and calculated the mean reddening $E(J-K_S)$ in distance bins to extract the primary trend of variation. We used the same color strategy in Fig. \ref{fig:warp} to draw the distance-mean reddening curve of the spatially-reclassified components in Fig. \ref{fig:bimodalE}. In principle, the two low-extinction groups allows a better chance to show ACS with a step in the reddening-distance plot because the contamination from the disc can be avoided there. However, we do not find significant difference beyond 2 kpc. The difference at around 1-2 kpc between these two low-extinction curves is probably due to the residual impact of the Local Arm and the Perseus Arm, instead of the Anti-Center Shell. 

\begin{figure}[h]
\centering
\includegraphics[scale=0.6]{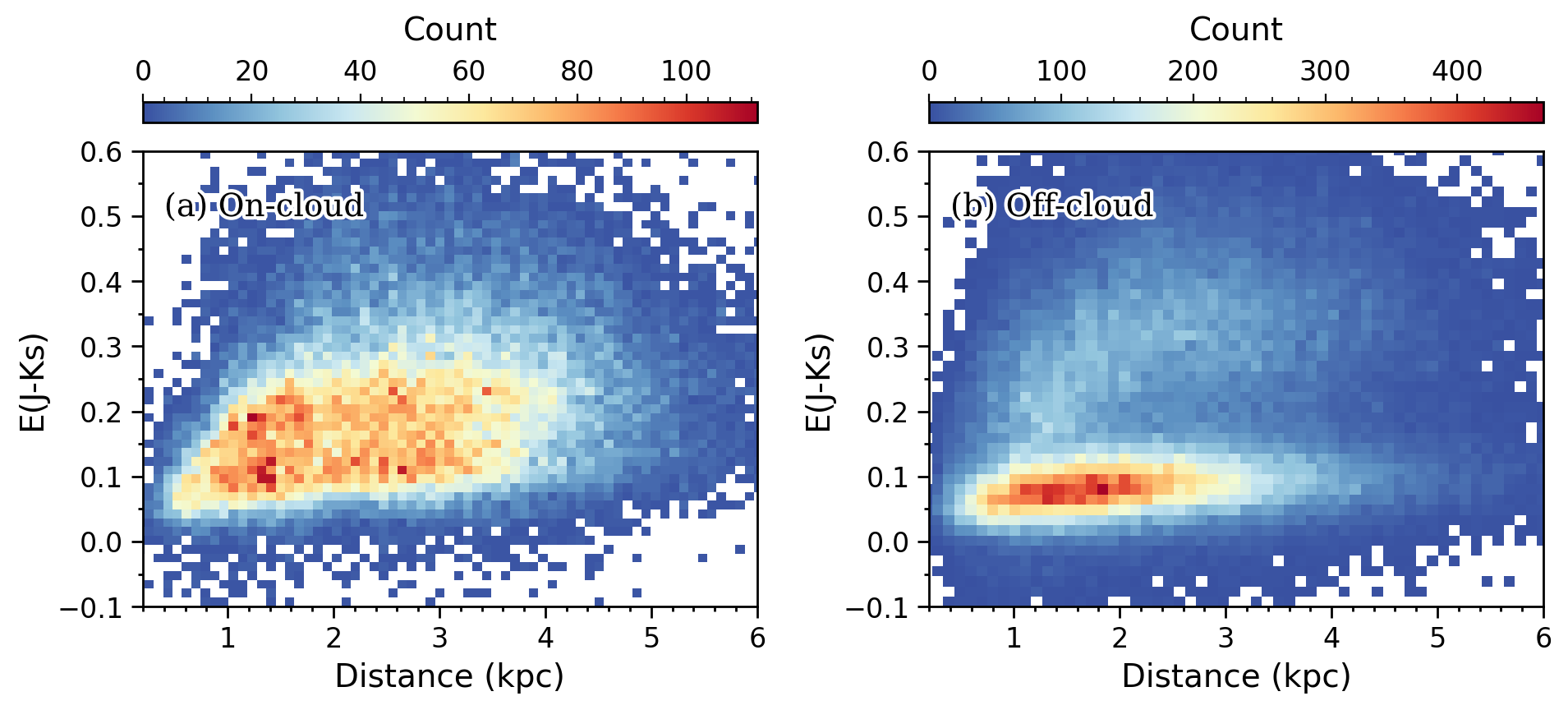}
\caption{Individual reddening of all K-giant stars (a) on-cloud; (b) off-cloud. There are two separated components in both plots: high-extinction one probably stemming from sight-lines going through the dusty Galactic disk, and low-extinction component going through higher-latitude low-density halo.} 
\label{fig:bi}
\end{figure}

\subsection{Negative velocity HV-DIBs and distance lower limit of Anti-center Shell (ACS)} \label{sec:res2}

With extinction and distance, we can only identify the step produced by ACS through comparisons of extinction variation in different regions. This method will be less effective when the signal is weak. However, we are able to obtain the velocity of DIB carriers at each point in the second method and screen out anomaly velocity incompatible with the Galactic disk. Since the Anti-Center Shell is a negative velocity HVC, we first inspected the negative velocity HV-DIBs in Fig. \ref{fig:HVDIBs} to see if they display any statistical structure in space. In Fig. \ref{fig:negHVDIB}, HV-DIBs with negative velocity do not show any clustering behavior at the same position with ACS' 21-cm \ion{H}{i} emission in three-dimensional space, which means there is no significant signal from ACS in the DIB measurements.

Non-detection of the cloud signal allows us to give the lower limit of ACS distance. We then dropped the stars with unreliable Bailor-Jones' distances whose signal-to-noise (SNR) values are smaller than one (note that we defined SNR here as \texttt{D/(D\_HIGH-D\_LOW)}). Eventually, we found the star with Bailor-Jones' distance $8.1 \pm 1.8$~kpc to provide the final distance lower limit.

\begin{figure}[h]
\centering
\includegraphics[scale=0.45]{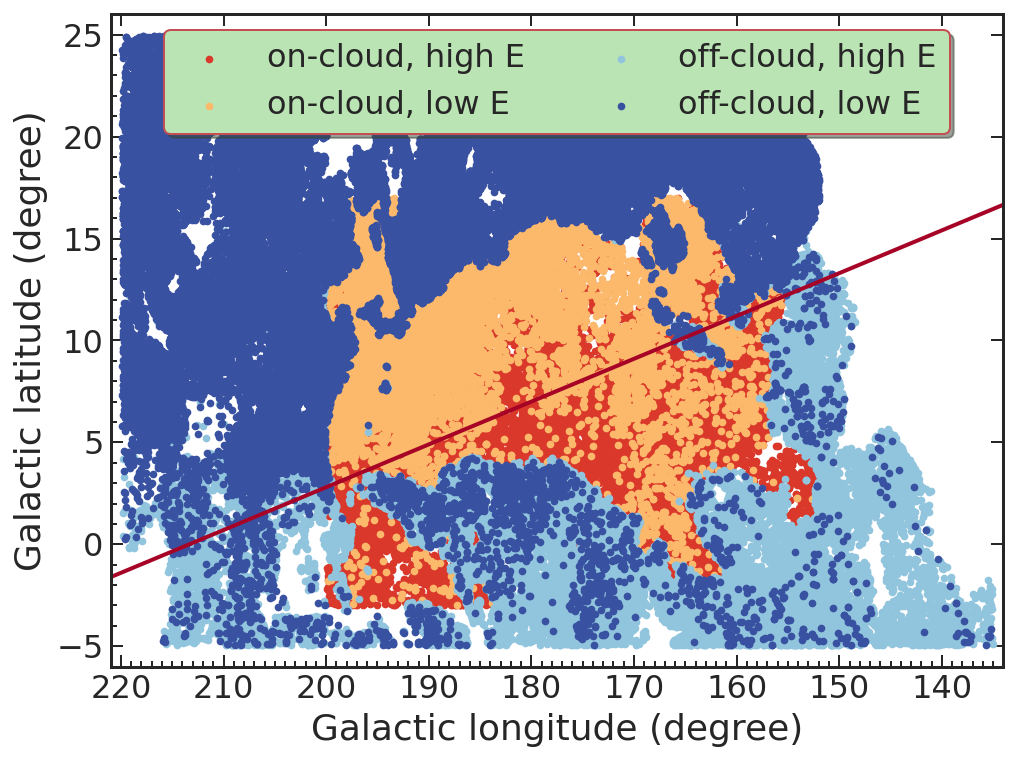}
\caption{Spatial distribution of different extinction modes. Red dots denote the high-extinction component on-cloud, orange ones denote the low-extinction component on-cloud, and light blue ones high-extinction component off-cloud, deep blue ones low-extinction component off-cloud. The red solid line shows the rough boundary among each component, which is consistent with the shape of the dust disk's warp of our Milky Way.} 
\label{fig:warp}
\end{figure}

\begin{figure}[h]
\centering
\includegraphics[scale=0.45]{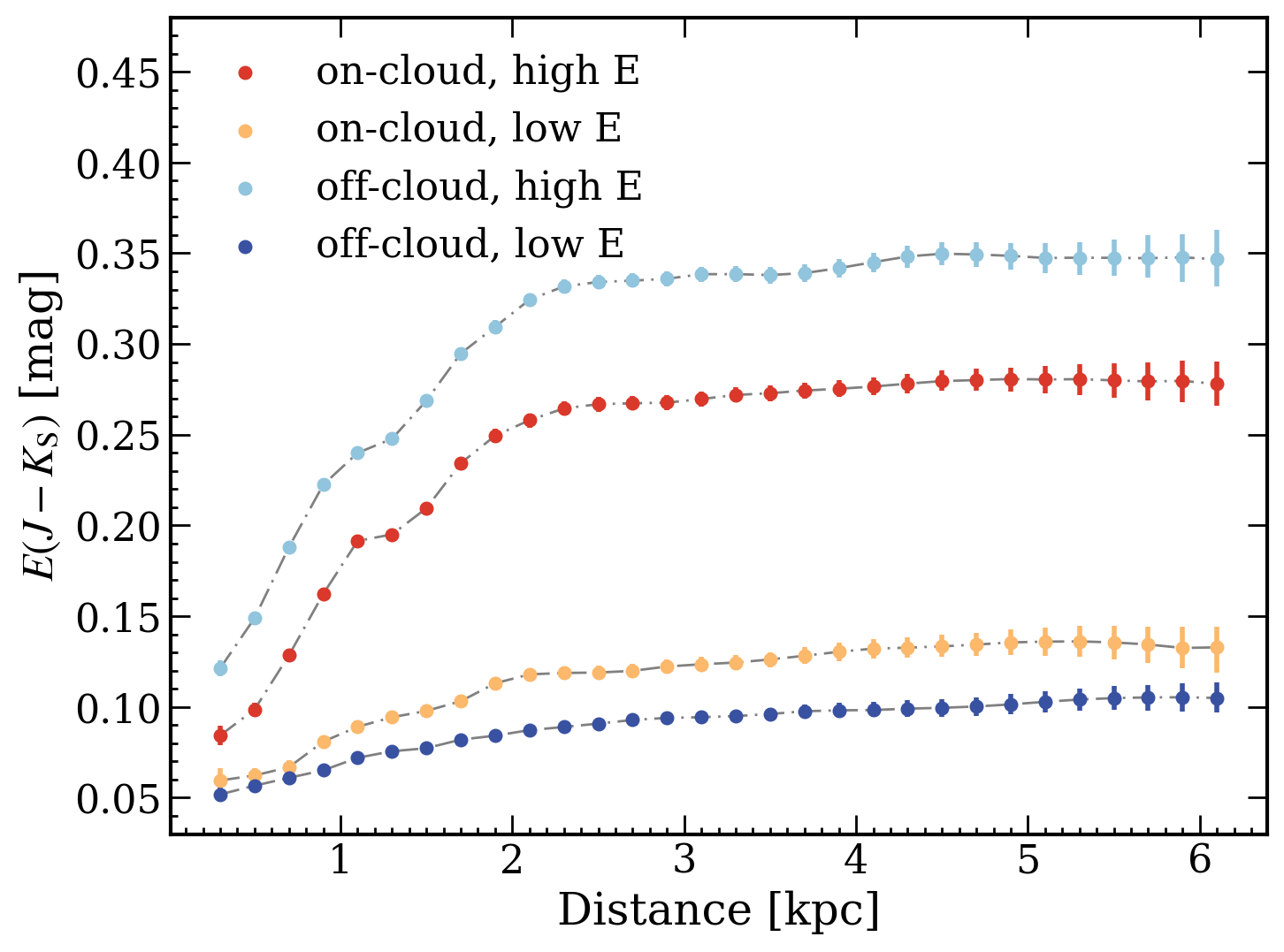}
\caption{Distance-average reddening relation for each extinction mode reclassified in spatial distribution. We use the same color strategy as that in Fig. \ref{fig:warp}.}
\label{fig:bimodalE}
\end{figure}

\begin{figure}[h]
    \centering
    \includegraphics[scale=0.38]{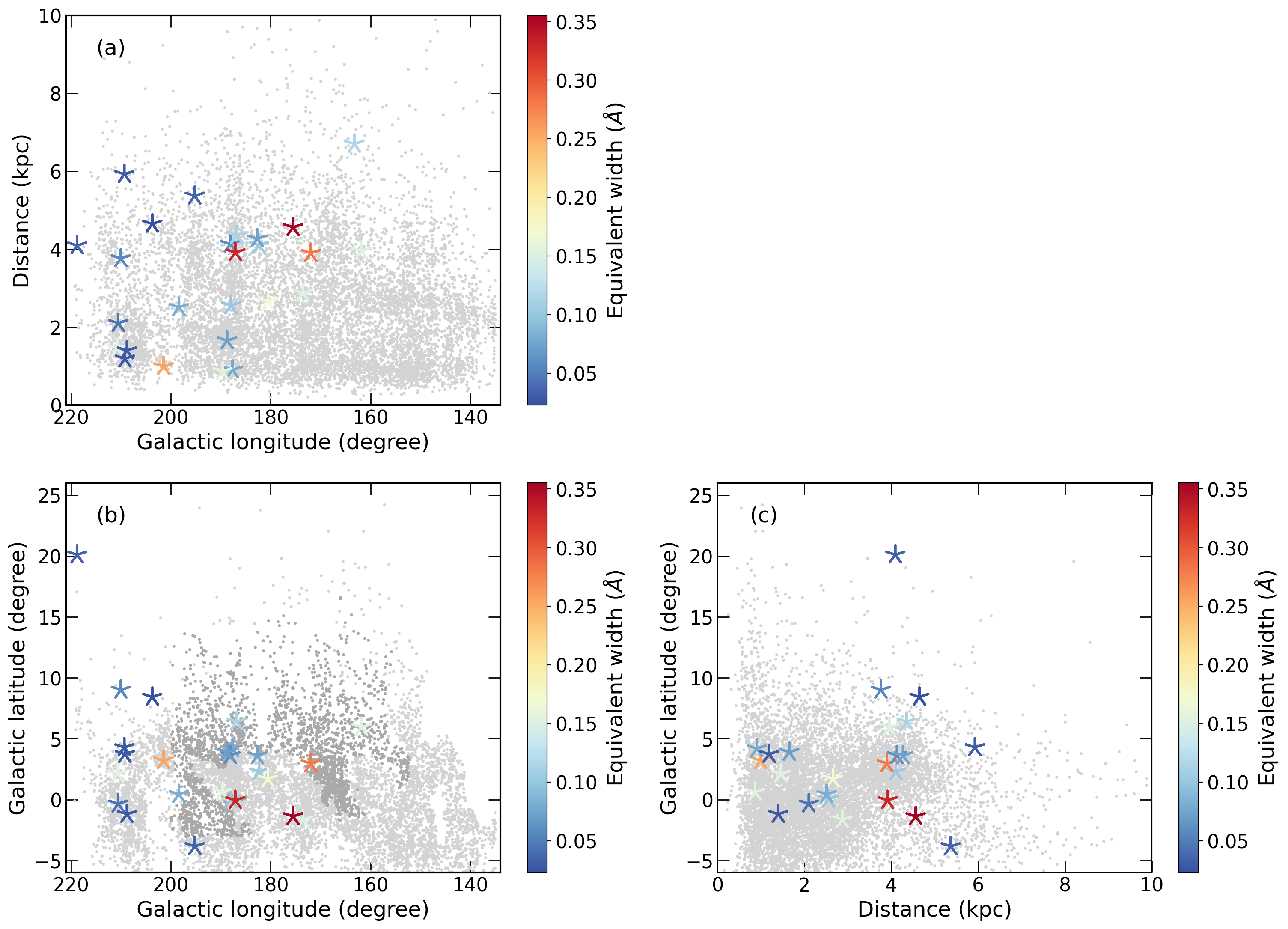}
    \caption{Three-dimensional positions of HV-DIBs with negative velocity. We used the distance of our O- and B-type star tracers accomplished by \citet{bailorjones2021-est}. HV-DIBs are color-coded by their equivalent width. Light gray dots on the background show all reliable DIB measurements, and dark gray dots in (b) accent the cloud area.} 
    \label{fig:negHVDIB}
\end{figure}

\section{Discussion} \label{sec:discussion}

\subsection{Stacked spectra for the Diffuse Interstellar band $\lambda$5780 in the cloud region} \label{sec:dis}

\begin{figure}[h]
    \centering
    \includegraphics[scale=0.6]{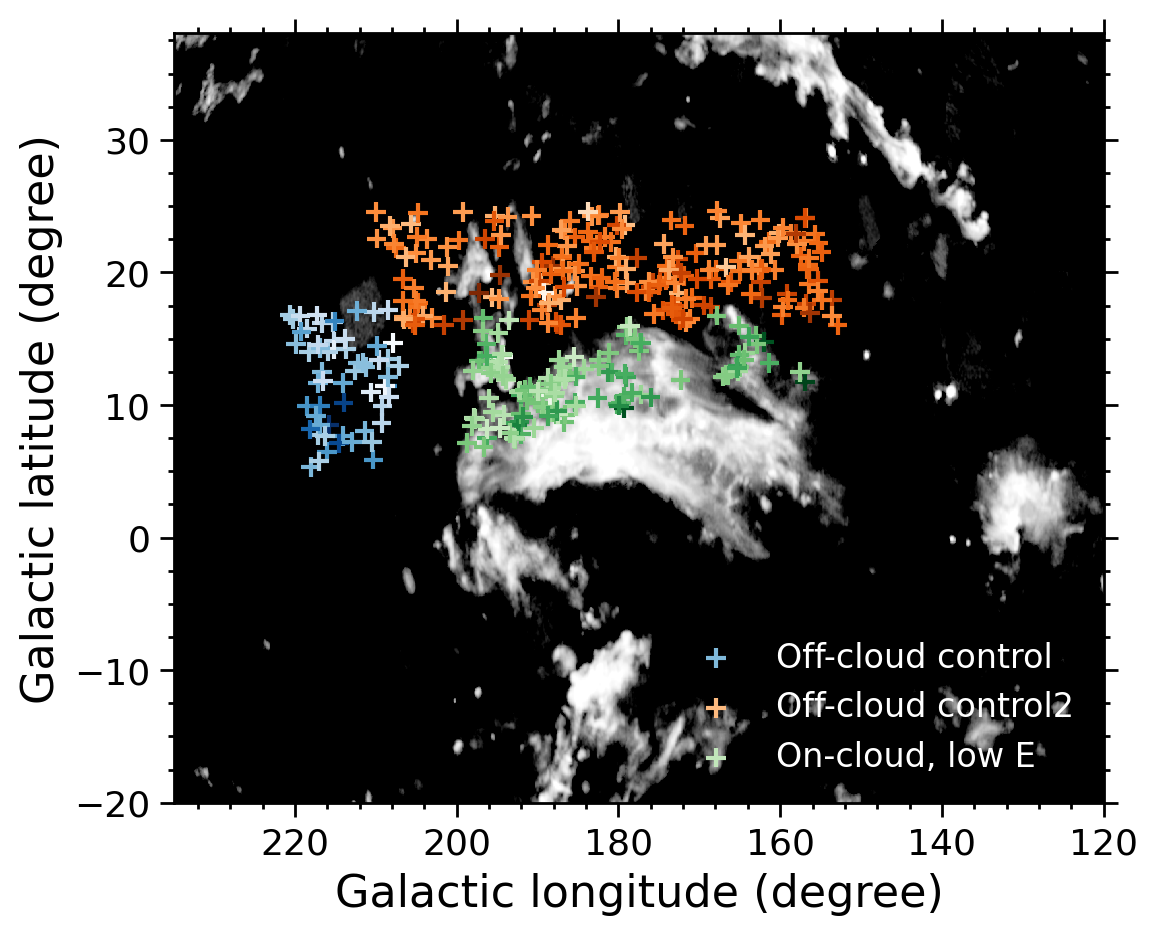}
    \caption{Illustration of the three regions used for stacking Diffuse Interstellar Band $\lambda$5780 spectra. Green crosses indicate O- and B-type stars projected onto the ACS cloud region, while blue and orange crosses denote stars in the off-cloud control regions. The color intensity of green, blue and orange represents the magnitude of reddening, respectively, for the stars in the three groups.} 
    \label{fig:stack_region}
\end{figure}

\begin{figure}[h]
    \centering
    \includegraphics[scale=0.6]{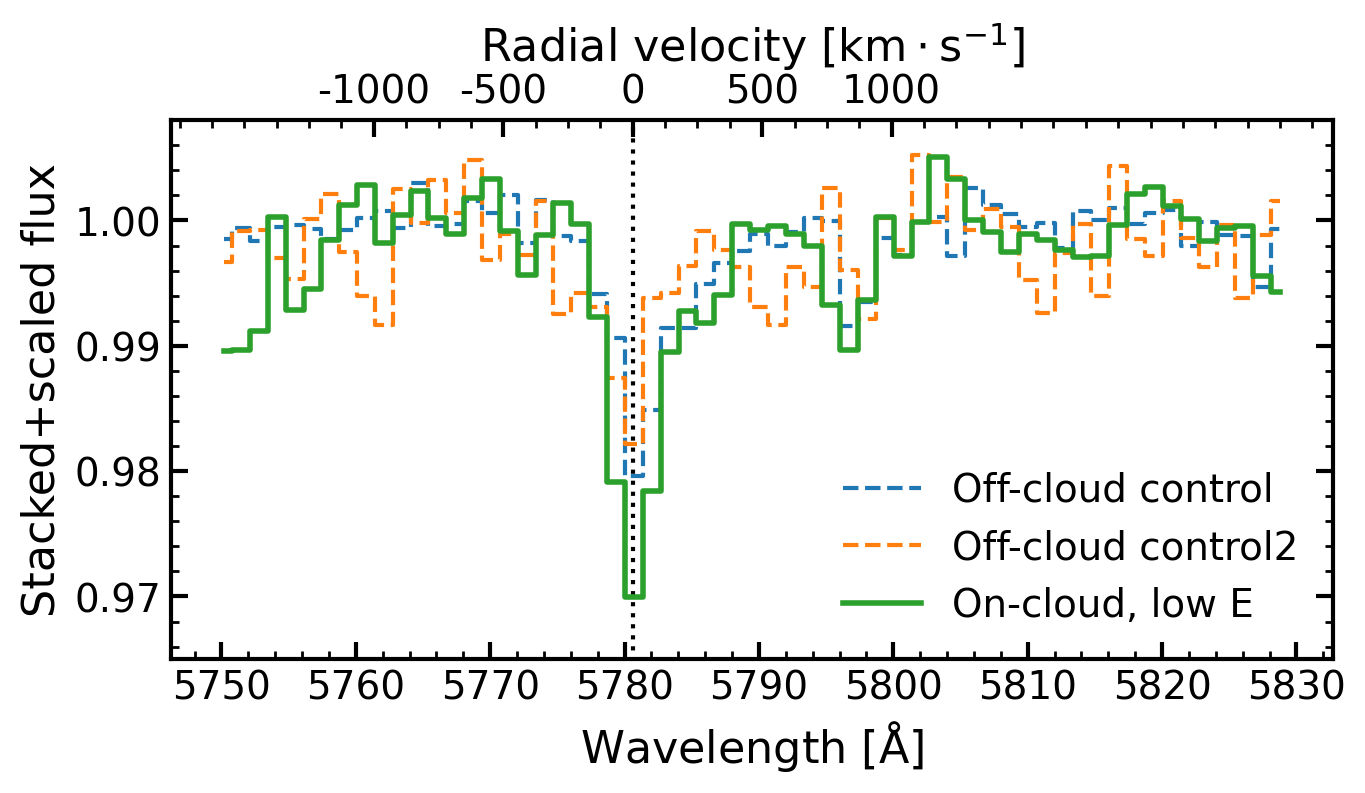}
    \caption{Stacked spectra of O- and B-type stars from three selected regions shown in Fig. \ref{fig:stack_region}. Each spectrum was normalized by its continuum prior to stacking. The green solid line represents the stacked spectra of stars in the cloud region, while blue and orange dashed lines show the stacked spectra from off-cloud control regions. The black dotted vertical line indicates the DIB $\lambda$5780 air wavelength of 5780.61~$\mathrm{\AA}$ \citep{hobbs2009}.} 
    \label{fig:stack_spec}
\end{figure}

In this section, we stacked the spectra of O- and B-type stars projected onto three selected regions (shown in Fig. \ref{fig:stack_region}) to enhance potential ACS signals. Prior to stacking, each spectrum was normalized by dividing by its continuum. The on-cloud experimental group, marked with green crosses in Fig. \ref{fig:stack_region}, excludes stars below the tilted line in Fig. \ref{fig:warp} to minimize disk contamination. Two off-cloud control groups, marked with blue and orange crosses, were selected at sufficient distances from the ACS cloud to avoid potential interference from its outer ionized regions. The continuum was calculated following the methodology described in Section \ref{sec:method2} \citep{zhaohe2021-diffuse}. Fig. \ref{fig:stack_spec} presents the results: the green solid line shows the stacked spectra of stars in the cloud region, while blue and orange dashed lines represent the stacked spectra from the off-cloud control regions.

As shown in Fig. \ref{fig:stack_spec}, the DIB $\lambda$5780 peaks align across all three regions, with no bimodal structure observed in the negative velocity (left-wing) portion of the absorption feature for the experimental group stars. While the DIB $\lambda$5797 feature in the on-cloud sample spectrum appears shifted toward more negative velocities, Gaussian profile fitting yields a velocity of $-32.6~\mathrm{km \cdot s^{-1}}$. This value is significantly different from the ACS velocity of $v_{\mathrm{LSR}}\sim -100~\mathrm{km \cdot s^{-1}}$. These results further support our non-detection of ACS using the DIB method.

\subsection{Explanations for non-detection of the Anti-Center Shell} \label{sec:dis1}

Non-detection of the the negative-velocity high-velocity cloud Anti-Center Shell in both the extinction method and the DIB method can be interpreted as distance limitation of our tracers, meaning that we can deduce the lower limit of the HVC in this scenario. Non-detection of the ACS by our methods can also caused by the sensitivity of our methods. The signal of ACS may be concealed by the strong component of the Galactic disk. Another explanation for our non-detection might be that the neutral hydrogen and dust in the ACS are not sufficient to cause remarkable extinction excess. We can see from Fig. \ref{fig:cloud} that the densest part of the cloud has a \ion{H}{i} column density of $\mathrm{log}_{10}(N_{\ion{H}{i}}) \sim 20$, and according to the $E(B-V)-N_{\ion{H}{i}}$ relation in Figure 11 of \citet{lan2015-ex}, this column density corresponds to negligible $E(B-V)$. Chances are that this cloud is also too hot to harbor dust and DIB carriers to produce significant enough extinction and cause the malfunction of our methods. This explanation is based on that the high-velocity cloud ACS' composition is similar to that of the all-sky cloud-averaged in the Milky Way, allowing us to apply the $N_{\mathrm{\ion{H}{i}}}-E(B-V)$ relation in \citet{lan2015-ex}, which is not necessarily true.

\subsection{Positive velocity HV-DIBs} \label{sec:dis2}

\begin{figure}[h]
\centering
\includegraphics[scale=0.38]{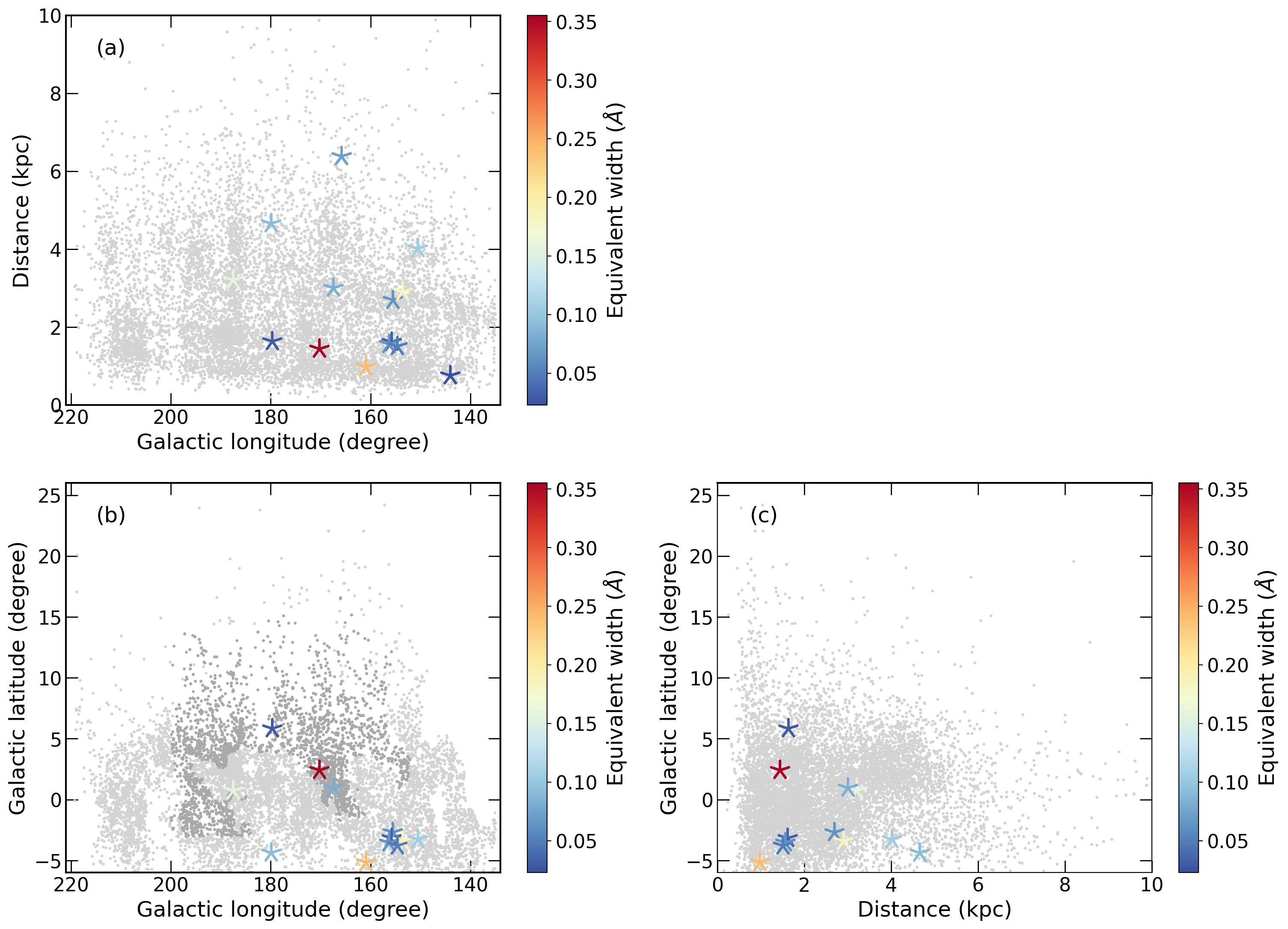}
\caption{Three-dimensional positions of HV-DIBs with positive velocity. We used the distance of our O- and B-type star tracers accomplished by \citet{bailorjones2021-est}. HV-DIBs are color-coded by their equivalent widths. Light gray dots on the background show all reliable DIB measurements, and dark gray dots in (b) accent the cloud area. DIBs at $(l,b)=(155,-5)^{\circ}$ show clustering characteristics in all three plots, clumpy-shaped in glon-glat plot and bar-shaped in plots relevant to distance.} 
\label{fig:posHVDIB}
\end{figure}

\begin{figure}[htb]
\centering
\includegraphics[scale=0.4]{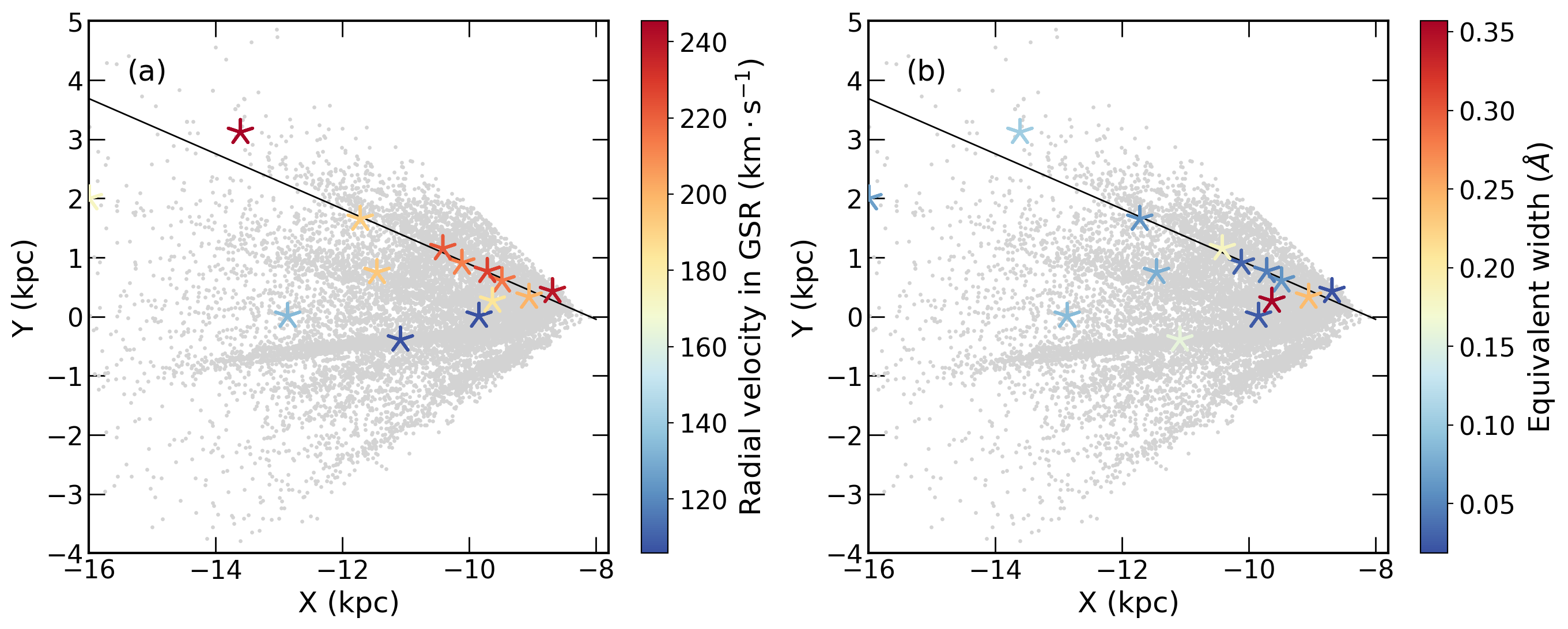}
\caption{High-quality positive velocity HV-DIBs (Asterisks color-coded in velocity and equivalent width) in X-Y plane. Gray dots denote all reliable DIB measurements. The angular position of the structure at $(l,b)=(155,-5)^{\circ}$ is emphasized with the black solid line.} 
\label{fig:xy}
\end{figure}

\begin{figure}[htb]
\centering
\includegraphics[scale=0.55]{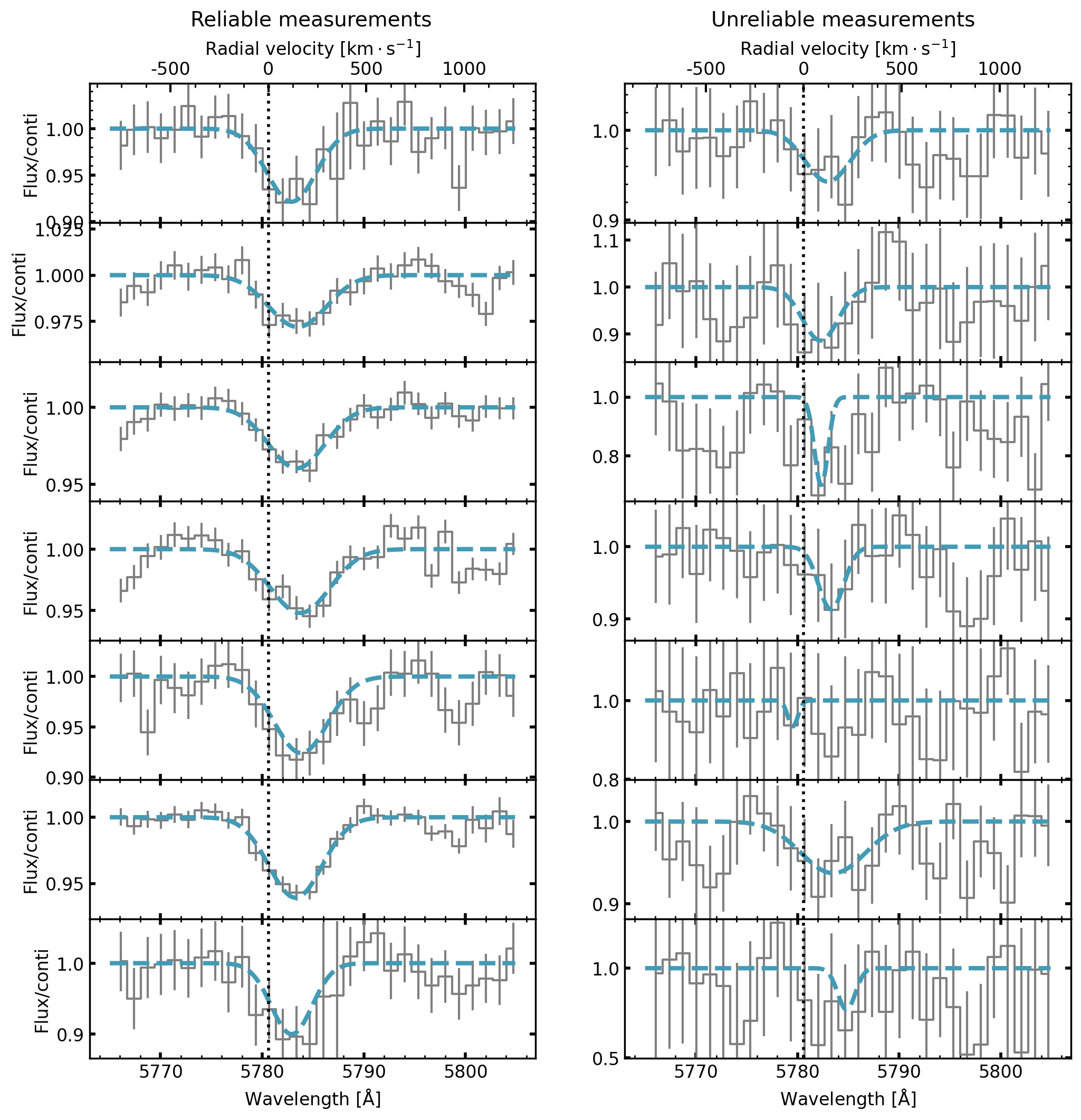}
\caption{Spectra and fitting results for stars with positive velocity HV-DIB detections. We mark the central air wavelength of DIB $\lambda$5780 in the rest frame with black dotted line, which is 5780.61 $\mathrm{\AA}$ measured by \citet{hobbs2009}. The Gray solid lines denote the observed spectra, while green dashed lines represent the Fitted Gaussian model. We performed an eye check and plotted some of the reliable measurements on the left side, and unreliable ones on the right side.} 
\label{fig:pos_spec}
\end{figure}

Though negative velocity HV-DIBs do not show any statistically clustering behavior, positive ones do present some intriguing clues. In Fig. \ref{fig:posHVDIB}, DIBs at $(l,b)=(155,-5)^{\circ}$ show clustering characteristics in all three plots, clumpy-shaped in glon-glat plot and bar-shaped in plots relevant to distance. We revisited our spectrum-fitting results of these positive velocity HV-DIBs and retain those good-fitting ones to draw in the X-Y plane in Fig. \ref{fig:xy} to see the integral variation in velocity and equivalent width along the strip-like structure. We plotted the eye check results of the spectrum fitting of stars with positive velocity HV-DIB detection in Fig. \ref{fig:pos_spec}. The reliable measurements are shown on the left side, while the unreliable ones on the right side.

Note that both line-of-sight velocity in GSR (Galactic Standard of Rest reference frame) and equivalent width are in a roughly increasing sequence along the pattern, which may indicate that the high-velocity DIB structure does have some extent of thickness along the $(l,b)=(155,-5)^{\circ}$ sight-line. The gradient of velocity in the structure is not very evident, ranging from about $200$ to $250~\rm{km} \cdot \rm{s}^{-1}$ in GSR when moving outward. Equivalent widths detected by stellar spectra down the stripe rise from about $0.3$ to $1.1$, meaning that the light of background stars with further distance goes through thicker DIB clouds or diffuse medium generating DIB features. However, this does not mean that this positive-velocity HV-DIB clump truly extended from 1~kpc to 5~kpc as we observe in Fig. \ref{fig:posHVDIB} since the asymmetric distribution of distance error of \textit{Gaia} can elongate the structure along the line of sight and cause the artificial shape of the cloud \citep{bailorjones2021-est}.

\section{Conclusion}\label{sec:conclusion}

In this paper, we have analyzed the distance of the high-velocity cloud Anti-Center Shell. We used interstellar extinction calculated on K-giant stars with LAMOST atmospheric parameters, 2MASS photometry and \textsl{Gaia} distance, and used Diffuse Interstellar Band feature at \qty{5780}{\angstrom} with LAMOST low-resolution spectra to support as our cloud tracers. 

In both methods, the signal of ACS is too weak to be found. In the extinction method, we compared the extinction variation of on- and off-cloud stars, but no excess extinction independently appearing in the on-cloud curve that we displayed. With the DIB method, it was possible for us to identify the cloud in velocity space. We should have discovered a component with an anomaly line-of-sight velocity in the LSR of about $-100~\rm{km} \cdot \rm{s}^{-1}$ if ACS was in the detectable range of our O- and B-type star tracers. But no prominent structure with negative velocity incompatible with the disk was uncovered in the Galactic longitude-Radial velocity plot.

Both analyses indicate that nowadays data is still incapable of detecting the ACS. The most intuitive explanation is that the cloud is even further than our tracers, which can lead us to give a distance lower limit of $\sim 8$~kpc. \citet{smoker2011-distance} also measured the distance to ACS using the absorption-line method. They owned two tracer stars possessing ultraviolet band spectra, HDE 248894 and HD 256725, but failed to identify the absorption by ACS as well and gave a lower limit of $8$~kpc. The distance of HDE 248894 and HD 256725 can be updated to $2.8$~kpc and $4.0$~kpc from Bailor-Jones' distance estimation based on \textit{Gaia} EDR3 \citep{bailorjones2021-est} now. So their lower limit should be altered to $\sim 4$~kpc, and our work pushes outward the lower limit to $\sim 8$~kpc, which means the Anti-Center Shell might have a Galactocentric distance of over $16$~kpc. Other explanations for our non-detection might be that: (1) the signal of the ACS can be concealed by the strong affection of the Galactic disk; (2) the neutral hydrogen and dust in the ACS are not sufficient to cause remarkable extinction excess.

The future direction of this study includes improving the number, spatial scale, and measurement quality (accuracy/resolution) of stellar distance, and gaining deeper stellar spectrum observation. Besides, we can study more about the physical nature of high-velocity clouds (e.g., their correlation with dust) to look for new detection methods.

\begin{acknowledgements}

The authors appreciate discussions with Jiadong Li.

We acknowledge the support of National Key R\&D Program of China No. 2019YFA0405500 and the China Manned Space Project with no. CMS-CSST-2021-A08.

We acknowledge the support of the National Natural Science Foundation of China (NSFC) under grants No. 12041305, 12173016.

We acknowledge the Program for Innovative Talents, Entrepreneur in Jiangsu. 

We acknowledge the science research grants from the China Manned Space Project with NOs.CMS-CSST-2021-A08 and CMS-CSST-2021-A07

This work used the data from the European Space Agency (ESA) mission Gaia (\url{https://www.cosmos.esa.int/gaia}), processed by the Gaia Data Processing and Analysis Consortium (DPAC; \url{https://www.cosmos.esa.int/web/gaia/dpac/consortium}). Funding for the DPAC has been provided by national institutions, in particular the institutions participating in the Gaia Multilateral Agreement. 

Guoshoujing Telescope (the Large Sky Area Multi-Object Fiber Spectroscopic Telescope LAMOST) is a National Major Scientific Project built by the Chinese Academy of Sciences. Funding for the project has been provided by the National Development and Reform Commission. LAMOST is operated and managed by the National Astronomical Observatories, Chinese Academy of Sciences.

\textsl{Facilities}: \textsl{Gaia}, LAMOST.

\textsl{Software}: \texttt{IPython} \citep{PER-GRA:2007}, \texttt{jupyter} \citep{2016ppap.book...87K}, \texttt{pandas} \citep{reback2020pandas}, \texttt{Astropy} \citep{2022ApJ...935..167A}, \texttt{numpy} \citep{harris2020array}, \texttt{scipy} \citep{2020SciPy-NMeth}, \texttt{matplotlib} \citep{Hunter:2007}.

\end{acknowledgements}

\appendix                  

\section{Uncertainty estimation of the extinction method} \label{sec:error_est}

\begin{figure}[h]
    \centering
    \includegraphics[scale=0.4]{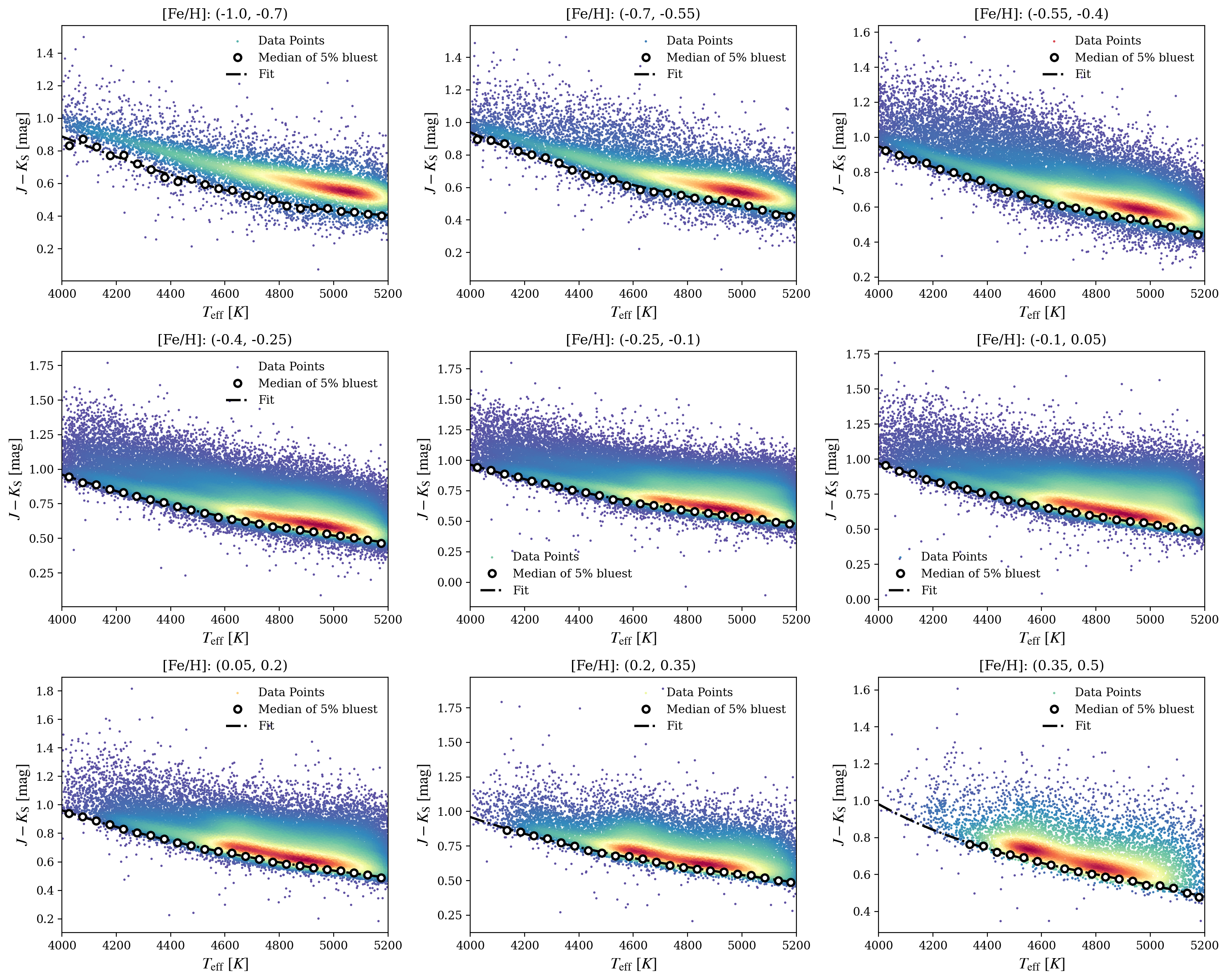}
    \caption{Blue edges calculated from the new iterative method for the metallicity bins [Fe/H] [-1.0, -0.7), [-0.7, -0.55), [-0.55, -0.4), [-0.25, -0.1), [-0.1, 0.05), [-0.05, 0.2), [0.2, 0.35), [0.35, 0.5]. Data points in different metallicity bins are plotted with color-coded solid dots, with color map denotes the Gaussian KDE density map of the points. The iteratively selected blue edge points are drawn with black circles. We also plot the polynomials fitted to the blue edge points with black dotted-dash lines.}
    \label{fig:blue_edges_RFR}
\end{figure}

\begin{figure}[h]
    \centering
    \includegraphics[scale=0.6]{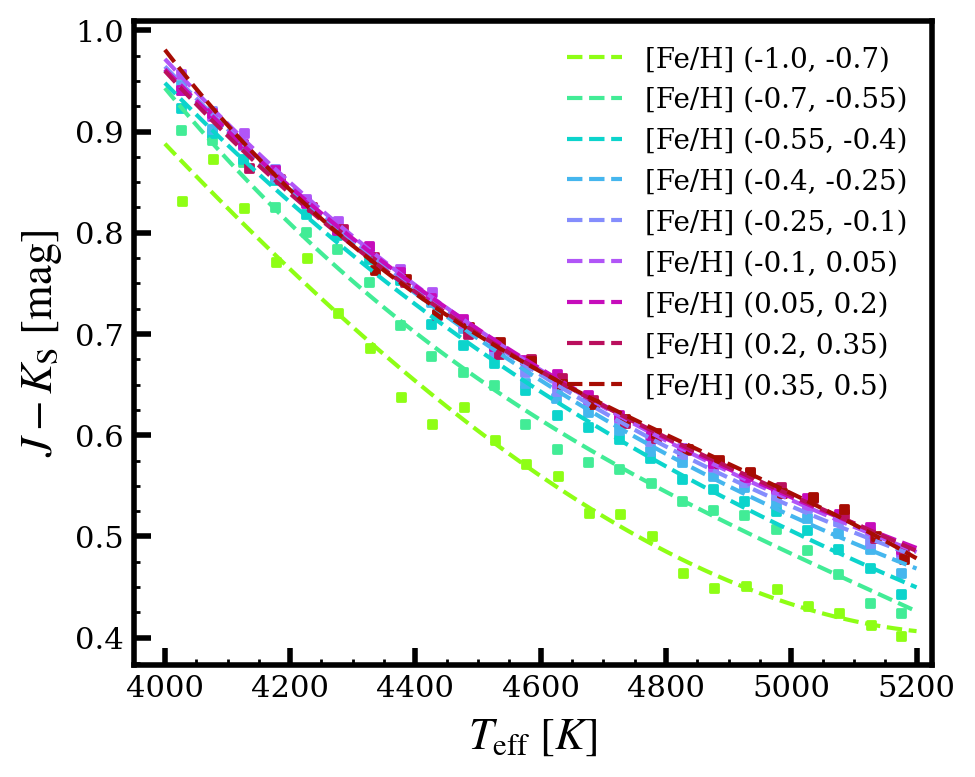}
    \caption{All the resulted blue edges in one plot. The square dots are the selected blue edge points, while the dashed lines denote the fitted polynomials to them.}
    \label{fig:blue_edges_con}
\end{figure}

\begin{figure}[h]
    \centering
    \includegraphics[scale=0.6]{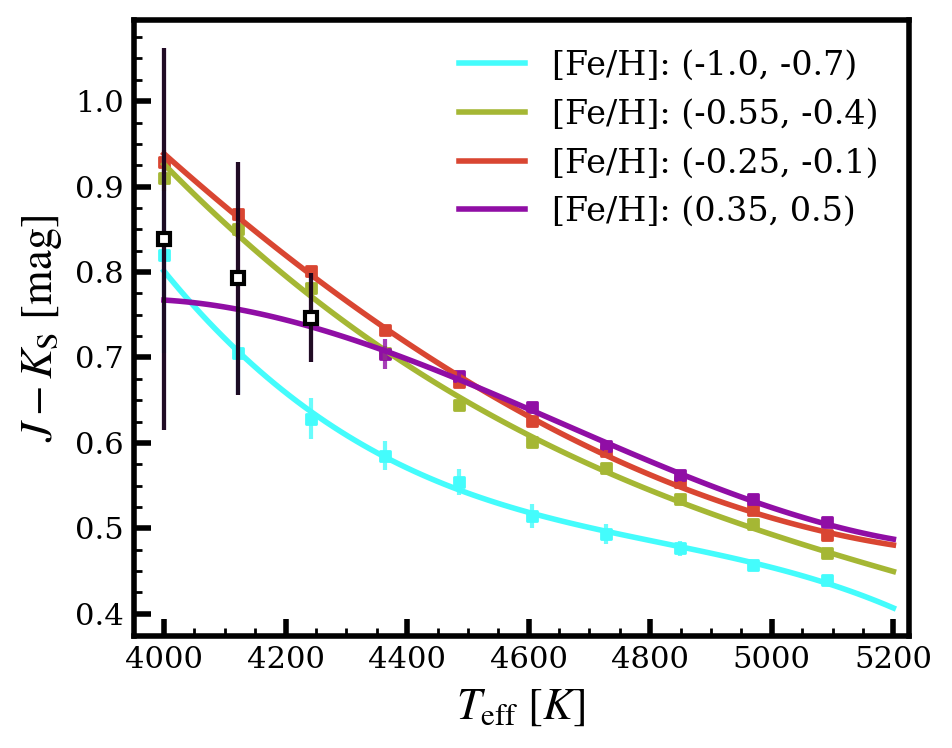}
    \caption{The Monte Carlo errors calculated for four different metallicity bins. In the formal determination of blue edge of the metallicity bin [0.35, 0.5], the blue edge points with $T_{\mathrm{eff}}<4350~K$ are discarded due to the small amount of data points.}
    \label{fig:blue_edges_err}
\end{figure}

\begin{figure}[h]
    \centering
    \includegraphics[scale=0.6]{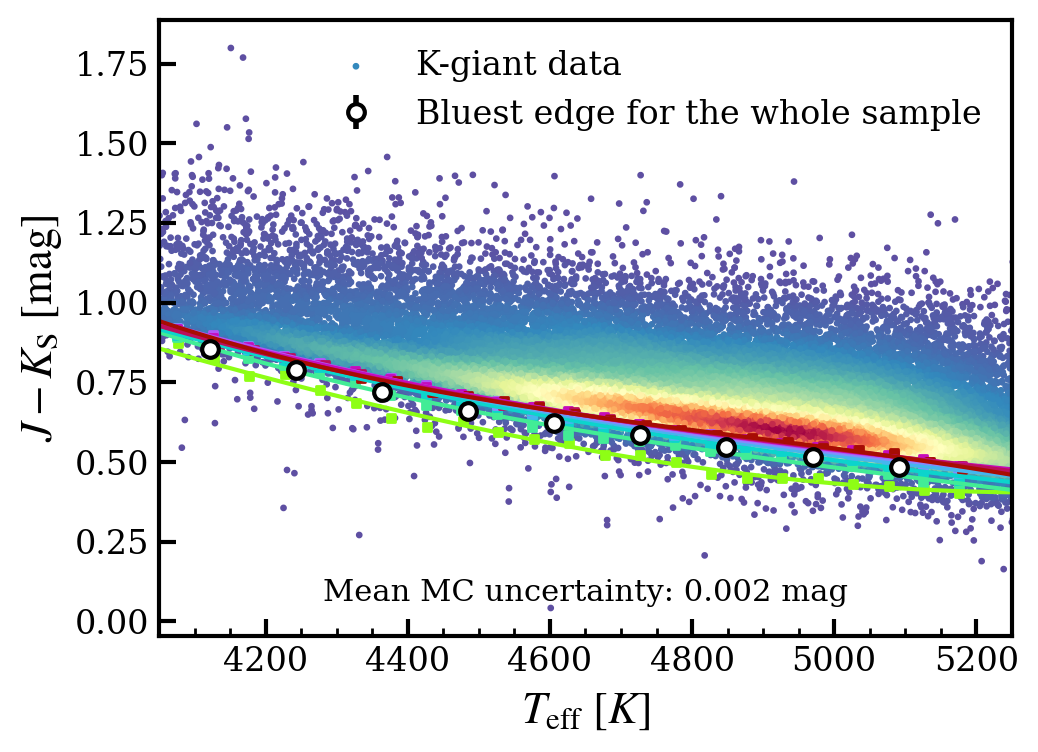}
    \caption{The blue edge point and Monte Carlo errors calculated for the whole sample with -1.0<[Fe/H]<0.5. The mean Monte Carlo uncertainty over the whole temperature range is $\sim$0.002 mag. The blue edges computed for different metallicity bins in Fig. \ref{fig:blue_edges_con} are also drawn for reference.}
    \label{fig:blue_edge_all}
\end{figure}

\begin{figure}[h]
    \centering
    \includegraphics[scale=0.6]{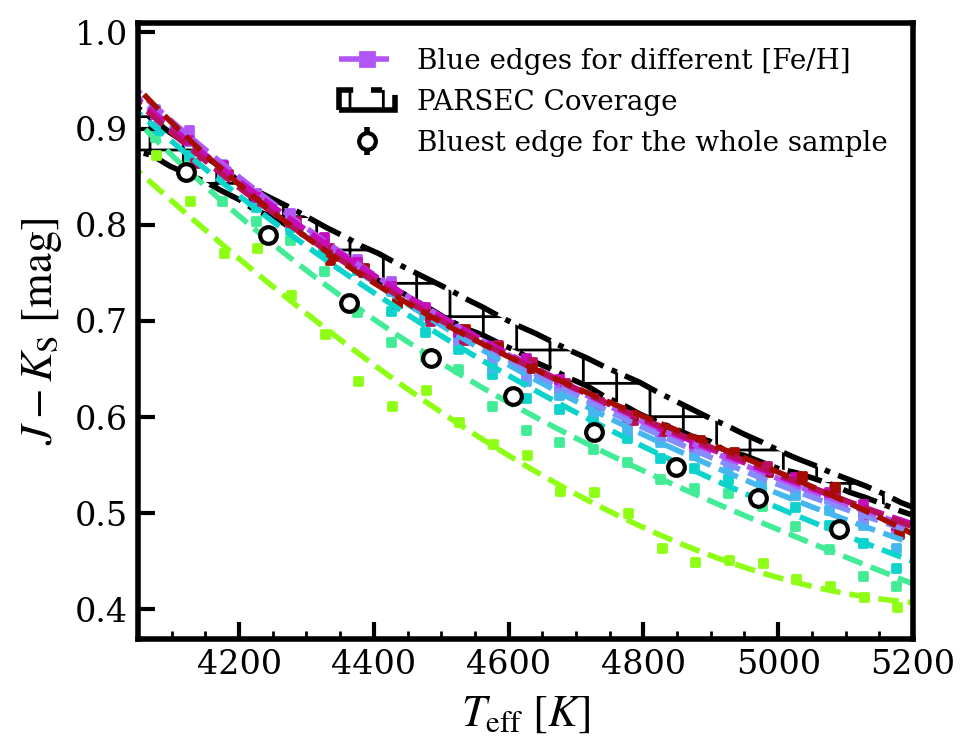}
    \caption{The dispersion of effective temperature – intrinsic color relation for RGB stars in PARSEC 1.2s. We select stars with logAge from 9.3 to 10.13, metallicity [M/H] from -1 to 0.5 in PARSEC. Meanwhile, we plot the blue edges computed for different metallicity bins in Fig. \ref{fig:blue_edges_con}, and the blue edge point and MC errors calculated for the whole sample with -1.0<[Fe/H]<0.5 for better comparison.}
    \label{fig:blue_edges_parsec}
\end{figure}

In this section, we will describe entirely how we estimate the uncertainties that might appear in the extinction method. The comprehensive uncertainty estimation includes the uncertainty of the bluest edge, the observational uncertainty of distance measurement, $J$-band and $K_{\mathrm{S}}$-band photometry, and the emerged Poison error when binning the data into sight line and distance bins.

\subsection{The uncertainty of the bluest edge}

The uncertainty of the blue edge is made up of systematic uncertainty arising from metallicity and age variations of the star sample, and observational uncertainty involved.

First, we examined the temperature-intrinsic color relation for K giants provided by the PARSEC models \citep{bressan2012,chen2014,chen2015,tang2014,marigo2017,Pastorelli2019,Pastorelli2020}. Using PARSEC 1.2s isochrones with ages ranging from 2 to 13.7 Gyr \citep[According to the SFH of Galactic disk from][]{nataf2024}, metallicities from -1 to 0.5, and a label of 3 (Red Giant Branch stage), we found that the broadening in the temperature-intrinsic color relation in the 2MASS $J$ and $K_{\mathrm{s}}$ bands due to age and metallicity variations is approximately 0.04 mag, which can be seen in Fig. \ref{fig:blue_edges_parsec}. This value gives a glimpse of the systematic error caused by the metallicity and age variation of K giants.

We also attempted to measure the bluest edge in different metallicity bins for our LAMOST DR5 K giants, using a modified version of the method outlined in \citet{jian2017} and \citet{cao2024}. We binned the stars by metallicity (width 0.15 dex, -1 to 0.5 dex; except for the bin of stars with poorest metallicity, for which we choose it to be -1.0 to -0.7 to guarantee enough stars) and effective temperature $T_{\mathrm{eff}}$  (width $\sim50~K$, excluding bins with fewer than 100 stars). To find the bluest edge, we iteratively selected the 5\% bluest stars ($J-K_{\mathrm{S}}$ color in 2MASS bands) in each temperature bin, fitted their color-temperature relation using Random Forest Regressor (RFR), and removed outliers exceeding 3 sigma uncertainty \citep[combining the photometric uncertainties in $J$ and $K_{\mathrm{S}}$ bands, and the uncertainty from LAMOST effective temperature, $\sim$0.030 mag according to][]{jian2017}. 
The obtained bluest edges for different metallicity bins are
plotted in Fig. \ref{fig:blue_edges_RFR} and \ref{fig:blue_edges_con} below. However, the observed variations in the bluest edge with metallicity were significantly larger than those predicted by the PARSEC models above, which is $\sim$0.1 mag versus $\sim$0.04 mag. According to \citet{jian2017}, this discrepancy likely additionally arises from the choice of fraction of the bluest stars (the method), Poisson noise, and observational errors in photometry, temperature and metallicity. \citet{jian2017} estimated that the different bluest fraction will introduce an error of about 0.02 mag.

So, we further estimated the uncertainty of the blue edges contributed by Poisson noise and observational errors in temperature and metallicity with the Monte Carlo method. The photometric and $T_{\mathrm{eff}}$ value of the bluest fraction are re-calculated 1000 times by the Monte Carlo method: each new value is the sum of the catalog value and a Gaussian random number determined by its observational error. We selected the metallicity bins [-1.0, 0.7), [-0.55, -0.4), [-0.25, -0.1) and [0.35, 0.5) to represent a quick estimate of the uncertainty provided by Poisson noise and observational errors over different metallicity ranges. We obtained that the blue edge uncertainties of the metallicity bins [-1.0, 0.7), [-0.55, -0.4), [-0.25, -0.1) and [0.35, 0.5) are 0.0198, 0.0045, 0.0029 and 0.0051 mag on average of the whole $T_{\mathrm{eff}}$ range, respectively. Noted that when we calculated the average uncertainty for the [Fe/H] bin [0.35, 0.5), we dropped the points with $T_{\mathrm{eff}}<4350~K$, for the star counts in those $T_{\mathrm{eff}}$ bins are smaller than 100. Bluest edges for the metallicity bins [-1.0, 0.7), [-0.55, -0.4), [-0.25, -0.1) and [0.35, 0.5) with Monte Carlo uncertainties are plotted in Fig. \ref{fig:blue_edges_err}. And for the whole sample with metallicity falling in range [-1.0, 0.5], its Monte Carlo uncertainty is plotted in Fig. \ref{fig:blue_edge_all}, which is only about 0.002 mag.

Overall, the systematic error caused by the different bluest fraction is ~0.02 mag \citep[according to][]{jian2017}, and the systematic error caused by the variation of age and metallicity is $\sim$0.04 mag (according to PARSEC). For the uncertainties caused by the Poisson error and observational errors, it will range from $\sim$0.003 to $\sim$0.020 mag if we bin the metallicity, and it will be $\sim$0.002 mag if we use the whole sample. In the main text, we applied the bluest edge calculated by the whole sample with [Fe/H] ranging from -1.0 to 0.5, so the overall uncertainty of the bluest edge is $\sim$0.045 mag.

\subsection{The uncertainty caused by the observation and the Poisson error}

For the observational uncertainty of distance measurement, $J$-band and $K_{\mathrm{S}}$-band photometry, we can account for them with the Monte Carlo technique to transfer the uncertainties of the observables to the final distance-extinction relation. 
The number of stars N in each distance and sight line bin also has Poisson statistical fluctuations, leading to additional uncertainty in the mean extinction. However, in each Monte Carlo simulation, stars may be assigned to different bins due to those observational errors, and this change in assignment indirectly reflects the statistical fluctuation of the number of stars in each bin (Poisson error). Therefore, we do not need to calculate it separately, it is already included in the simulation process. Similarly as in the last section, the distance, photometric and $T_{\mathrm{eff}}$ value of the bluest fraction are re-calculated 1000 times by the Monte Carlo method: each new value is the sum of the catalog value and a Gaussian random number determined by its observational error. Then we can obtain 1000 realization of the distance-extinction relation, and determine their 16, 50, 84 percentiles. The overall uncertainty can be calculated for each distance and sight-line bin as:

\begin{equation}
\sigma_{\mathrm{total}}=\sqrt{\sigma_{\mathrm{BluestEdge}}^2+\sigma_{\mathrm{MonteCarlo}}}
\label{eq:error}
\end{equation}

\bibliographystyle{raa}
\bibliography{ms2024-0390}

\end{document}